\documentclass[reprint,superscriptaddress,aps,prl]{revtex4-1}
\usepackage{graphicx}
\usepackage{xcolor,color}
\usepackage{amsmath} 
\usepackage{amsthm} 
\usepackage{amssymb}	
\usepackage{hyperref}
\newcommand{\cev}[1]{\overleftarrow{#1}}

\newcommand{\uv}[1]{\ensuremath{\mathbf{\hat{#1}}}} 
 
\newcommand{\ket}[1]{\left| #1 \right>} 
\newcommand{\bra}[1]{\left< #1 \right|} 
\newcommand{\braket}[2]{\left< #1 \vphantom{#2} \right|
	\left. #2 \vphantom{#1} \right>} 
\newcommand{\matrixel}[3]{\left< #1 \vphantom{#2#3} \right|
	#2 \left| #3 \vphantom{#1#2} \right>} 
\let\baraccent=\= 

\begin{document}

\title{Computational Framework for Angle-Resolved Photoemission Spectroscopy}

\author{R. P.\,Day}
\email[]{rpday7@gmail.com}
\author{B. \,Zwartsenberg}
\author{I. S.\,Elfimov}
\author{A.\,Damascelli}
\email[]{damascelli@physics.ubc.ca}
\affiliation{Department of Physics $\&$ Astronomy, University of British Columbia, Vancouver, BC V6T 1Z1, Canada}
\affiliation{Quantum Matter Institute, University of British Columbia, Vancouver, BC V6T 1Z4, Canada}

\date{\today}

\begin{abstract}

We have developed the numerical software package $chinook$, designed for the simulation of photoemission matrix elements\footnote{The full $chinook$ software package and documentation are available at \url{https://www.github.com/rpday/chinook}}. This quantity encodes a depth of information regarding the orbital structure of the underlying wavefunctions from which photoemission occurs. Extraction of this information is often nontrivial, owing to the influence of the experimental geometry and photoelectron interference, precluding straightforward solutions. The $chinook$ code has been designed to simulate and predict the ARPES intensity measured for arbitrary experimental configuration, including photon-energy, polarization and spin-projection, as well as consideration of both surface-projected slab and bulk models. This framework then facilitates an efficient interpretation of the photoemission experiment, allowing for a deeper understanding of the electronic structure in addition to the design of new experiments which leverage the matrix element effects towards the objective of selective photoemission from states of particular interest. 
\end{abstract}
\maketitle

\section{Introduction}

Angle-resolved photoemission spectroscopy (ARPES) and its variants have developed in recent years to be established among the pre-eminent experimental methods in solid-state physics. With an intimate connection to the one-electron removal spectral function, ARPES is unique among the suite of techniques available to condensed matter physicists in its direct correspondence to the electronic structure of crystalline materials, providing access to the one electron removal spectral function within its native momentum space \cite{Mahan,Hufner,scripta}.

In the framework of Fermi's Golden Rule, the photoemission intensity is described as:
\begin{equation}\label{eq:eq_fgr}
I(k,\omega) \propto \sum_{i,f} A_{f,i}(k,\omega)|\bra{\psi_f}\uv{\Delta}\ket{\psi_i}|^2,
\end{equation}

where $A_{f,i}$ is the one-electron removal spectral function, and $|\bra{\psi_f}\uv{\Delta}\ket{\psi_i}|^2$ the photoemission matrix element. The spectral function:

\begin{equation}
A_{f,i}(k,\omega)=\matrixel{\Psi_f^{N-1}}{c_k}{\Psi_i^{N}}\delta(\omega-E^{N-1}_f+E^{N}_i ),
\end{equation}
reflects the overlap between the initial $N$-particle many-body wavefunction upon removal of an electron and the ensemble of $(N-1)$-particle final state wavefunctions, while preserving energy conservation. Written as the imaginary part of the retarded Green's function:
\begin{equation}\label{eq:spectral}
A(k,\omega)=\frac{1}{\pi}\frac{-\Sigma"(k,\omega)}{(\epsilon_k^0 -\omega - \Sigma'(k,\omega))^2 + \Sigma"(k,\omega)^2}.
\end{equation}

The spectral function is seen to carry details of both the underlying bare dispersion associated with the electronic structure of the material $\epsilon_k^0$, as well as correlations via the self energy $\Sigma(k,\omega) = \Sigma'(k,\omega) + i \Sigma"(k,\omega)$. In the opposing limits of vanishing and strong interactions, ARPES is described as an ideal probe of the bandstructure and correlation effects respectively. 

In practice, the photoemission can be strongly modulated by the $|\bra{\psi_f}\uv{\Delta}\ket{\psi_i}|^2$ term, altering the spectral intensity through the dependence of the initial and final states on energy, momentum, and band index. At worst, this suppresses all intensity from certain bands, precluding their study by ARPES entirely. From a different perspective however, this modulation can be viewed as an additional experimental signature in the ARPES intensity which encodes a description of the photoemitted electron's wavefunction. 

This term can be simulated to allow for quantitative descriptions and insights regarding the experimental signal. While such an approach has been made at some level for a number of ARPES experiments, this requires substantial effort in developing a specific model for each study \cite{Bansil,shen,Gierz,Chiang,Jason,Gotlieb}. The development of a standard numerical framework would allow for a much larger set of experiments to be analysed at this level, providing the opportunity to understand and leverage the matrix element effects in a broad class of materials. We have pursued this objective through the development of an open source software package, $chinook$, implemented in Python to enable a broad audience to perform quick and easy simulation of photoemission-related phenomena, thereby improving both the interpretation and analysis of experimental data. 

In the following, we will outline the primary workflow of our numerical approach, and the various ways in which this package can be applied to the study of the electronic structure of solids via ARPES. 
\section{Results}
\subsection{Matrix Element Effects}

\begin{figure}[t!]
\centering
\includegraphics[width=\columnwidth]{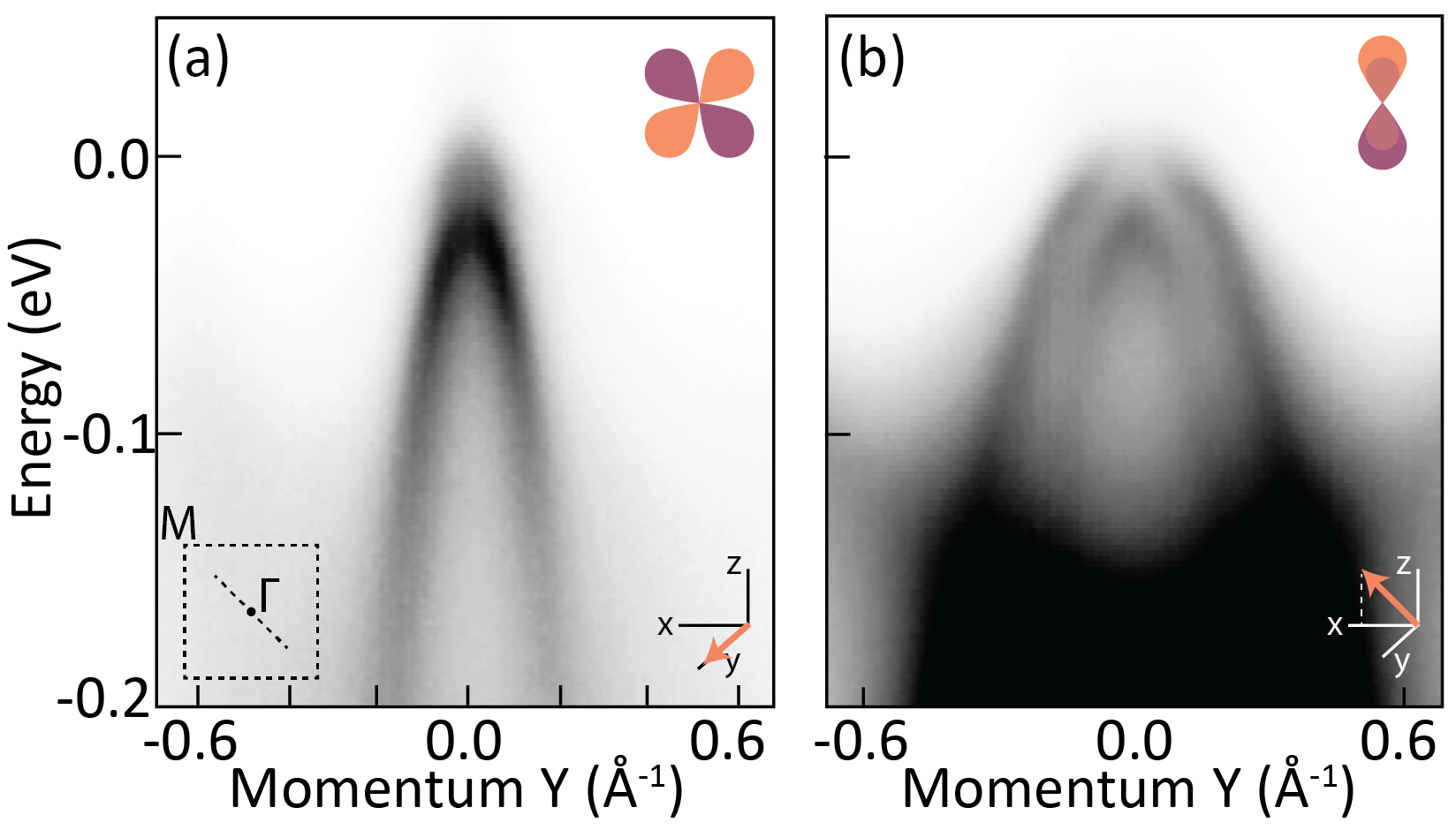}
\caption{Experimental ARPES on FeSe. Both panels display ARPES intensity from valence states taken at h$\nu=37$ eV and 120 K, directed along the $\Gamma M$ direction. Polarization is set to linear vertical (a) and horizontal (b), allowing for photoemission from states of different orbital character, as indicated by the insets. Adapted from Ref. \cite{Day}.}
\label{Fig:Fig1}
\end{figure}

In the design of an ARPES experiment, a cursory understanding or prediction of the matrix elements relevant to a given system can dramatically improve one's ability to study aspects of the electronic structure of interest. Before proceeding to explicit description of our software, it is instructive to consider a motivating example, taken here to be the iron-based superconductor FeSe. In Fig. 1, we plot experimental data along the $\Gamma M$ direction taken with two, orthogonal linear polarizations of light at $h\nu=37$ eV.  Near the Brillouin zone centre, three hole-bands disperse away from the Fermi level (c.f. Fig. \ref{Fig:Fig2}). For light polarized along the momentum axis in Fig. \ref{Fig:Fig1}a, only a single state is observed clearly, whereas the perpendicular polarization in Fig. \ref{Fig:Fig1}b. illuminates this and several other states. The third hole band is almost imperceptible, for any choice of polarization. These observations can be explained through an understanding of the orbital structure of the underlying electronic states, indicated by the insets of Fig. \ref{Fig:Fig1}, in combination with the orbital-mixing effects of spin-orbit coupling (SOC), as will be discussed in more detail below \cite{watson,Day}.  Recent experiments designed with these effects in mind have leveraged the dipole matrix elements to perform targeted spin- and angle- resolved photoemission from states of particular interest, extracting fundamental information pertaining to a broad variety of orbital-related phenomena \cite{dessau,veenstra,CD_TI,wathub,Day,Berend}. 

In many cases, orbital symmetry can be extracted from the polarization dependence alone. The information available from these arguments is limited, in particular for high-orbital ($l>1$) states, where the dimension of the vector field provides insufficient means to  identify all orbitals uniquely. This is further complicated in multi-atom bases, where now relative phases between different sites can differ from symmetric combinations in a momentum-dependent fashion. In these situations, the photoemission intensity pattern is found to depend sensitively on the relative phases within the initial state wavefunction, producing so-called photoelectron interference patterns. In this way however, the matrix elements encode further information regarding the initial state beyond orbital symmetry alone. These effects have been seen in for example graphene \cite{Chiang,Marchenko} and topological insulators \cite{Jason, Gotlieb}, demonstrating the full depth of information regarding the initial state wavefunction which is contained in the ARPES matrix element. To further leverage the information available from ARPES experiments, it is advantageous to be able to simulate the full ARPES experimental intensity, while maintaining physical transparency. By preserving access to the relevant model parameters, one can then establish a more fundamental, and conceptual understanding of the electronic structure.
 
\subsection{Model Hamiltonian}
There are various levels at which the ARPES matrix element can be modeled \cite{Mahan,Grobman,Bansil,Gierz,moser,Kruger}. While the most sophisticated approaches account for the possibility of scattering processes subsequent to the photoemission event such as those which make use of Korringa-Kohn-Rostoker final states \cite{mulazzi,minar,CuKKR}, we make two important simplifying assumptions here. First, the final states are taken to be free-electron plane waves:
\begin{equation}\label{Eq:Rayleigh}
\ket{e^{i\vec{k}\cdot\vec{r}}} = \sum_{l,m} i^l j_l(kr) Y_l^m(\theta,\phi) Y_l^{m*}(\theta_k,\phi_k).
\end{equation}
At high photon energies, the assumption of the plane wave final state is particularly well justified, as the crystal potential can be treated as a perturbation and sensitivity to the momentum structure of the exact final states becomes negligible \cite{scripta}. The validity of this assumption is ultimately material dependent, however similar logic as that applied to the  domain of suitability for the Born approximation can be made: such an assumption is reasonable when either the crystal potential $V_o\ll \hbar^2/m_e a^2$ or in the high-energy limit, $V_o/(\hbar^2/m_ea^2)\ll ka$, where $a$ is the range of the potential. At present it is possible within $chinook$ to relax this assumption only in the restricted sense of Ref. \onlinecite{Grobman}, as one can include phase shifts to the final state expansion. While beyond the scope of $chinook$ in its current form, it would be possible to write the final states in the form of more sophisticated scattering final states, where  the radial and orbital components of the ket in Eq. \ref{Eq:Rayleigh} are modified appropriately to reflect the presence of a finite crystal potential. This can be done through modification of the radial integrals $B_b^{l'}$ defined below.

Secondly, we work within a tight-binding framework wherein the initial states can be described by localized atomic-like orbitals, centred on the sites of the lattice basis. In materials where the spin degree of freedom is relevant, the orbital basis can be doubled to define a complete spinor basis, represented here by $\chi_{\pm}$. Formally, the tight-binding basis set is expressed typically as:

\begin{equation}\label{Eq:Eq_psi}
\phi_a = R^{a}_{n,l}(r)K^{a}_{l}(\Omega)\chi_{\pm},
\end{equation}
where $a$ represents a basis index and $n$, $l$ the principal and orbital quantum numbers respectively. $R^{a}_{n,l}(r)$ is a radial wavefunction, $K^{a}_{l}(\Omega)$ a cubic harmonic, and $\chi_{\pm}$ the spinor projection.  Alternatives such as distorted and rotated basis states can also be accommodated, so long as a unitary transformation into the basis of spherical harmonics can be made for the purpose of photoemission calculations. While these simplifications are in some cases unable to capture the full structure of the experimental photoemission intensity, we trade this level of universality for the substantial gains in transparency and physical insight which can be extracted from this approach. 

\begin{figure}[t!]
	\includegraphics[width=\columnwidth]{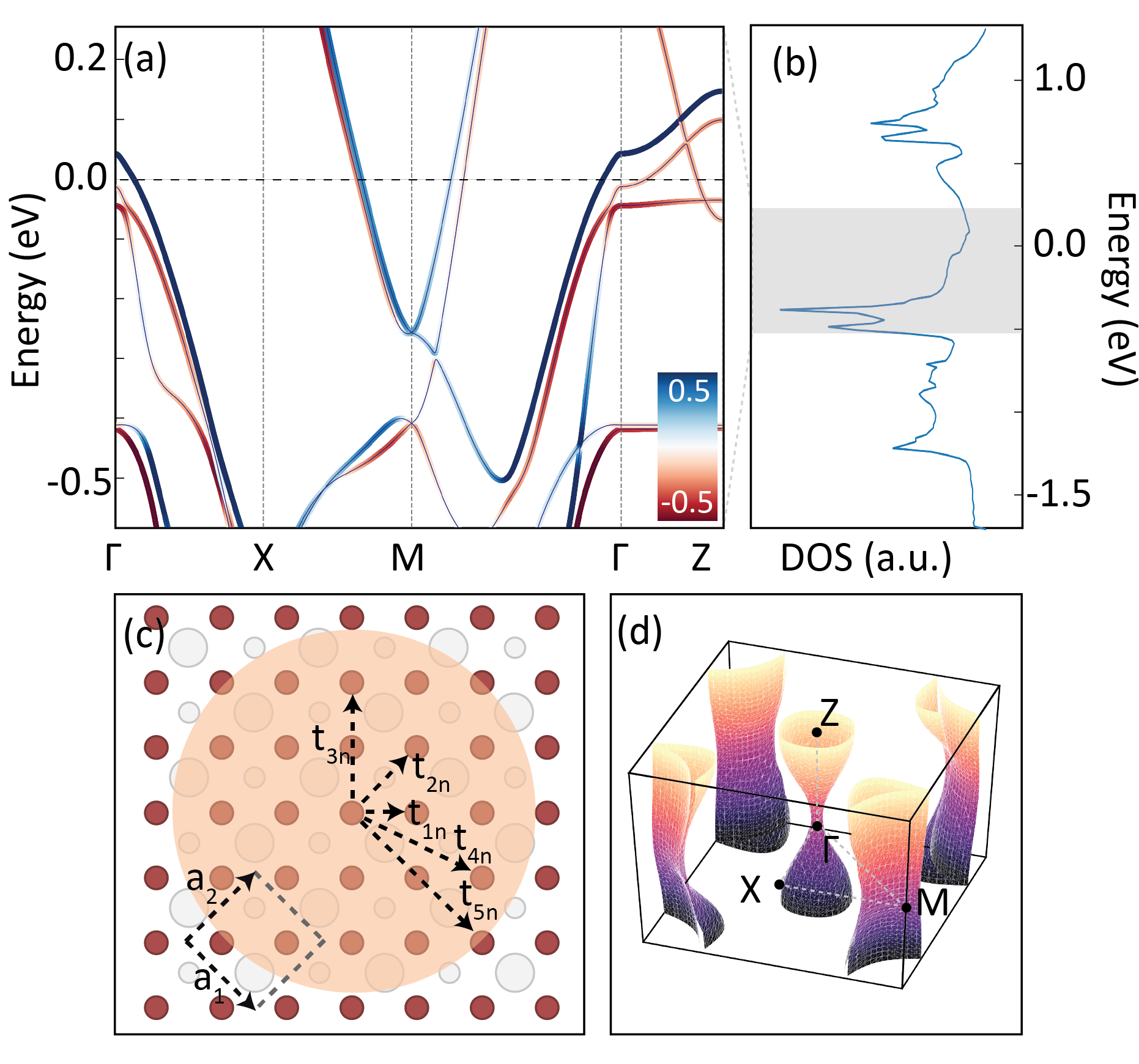}
	\caption{Tight-binding model of FeSe. Built using modified model from Ref. \onlinecite{EK_FeSC} for tetragonal FeSe. In (a) we plot the bandstructure along a high symmetry path, with the colourscale indicating the expectation value of $\left< \vec{L}\cdot\vec{S}\right>$. The Fe-3$d$ density of states is shown in (b). In (c) we plot the crystal structure projected into 2D, with Fe in red and the Se above (large) and below (small) in grey. In plane hopping terms are indicated, in addition to the primitive unit cell. The Fermi surface is plotted in (d).}
	\label{Fig:Fig2}
\end{figure}
Regarding the definition of the tight-binding model, there are various formalisms which are found in the literature, including Slater-Koster \cite{SK}, $t_{ab}$, and Wannier Hamiltonians: we have made an effort to accommodate all possible variations without loss of functionality. We require only that the Hamiltonian matrix elements can be written as a Fourier series of bilinear terms in the orbital Hilbert space:
\begin{equation}
H_{o}^{ab}(\vec{k}) = \sum_{\{\vec{r}_{ab}\}} t_{ab}e^{i\vec{k}\cdot\vec{r}_{ab}}c_{k,a}^{\dagger}c_{k,b}.
\end{equation}
In this expression $r_{ab}$ denotes the full connecting vector between basis states $\phi_a$ and $\phi_b$, as opposed to the equivalent form where one refers to the connecting lattice vector alone. In addition to $H_o(k)$, any other bilinear functions of momentum can also be added to the full Hamiltonian at this stage, including spin-orbit coupling, and orbital or spin order. Adherence to Bloch's theorem can be further relaxed in the context of the low-energy effective models which describe a narrow region of momentum space: in this scenario, a more general function of momentum which satisfies the point-group symmetry can be employed \cite{vafek}. 

With the basis and Hamiltonian so defined, the eigenvalue problem can be solved, and the initial state wavefunctions then defined as a superposition of the basis states described by Eq. \ref{Eq:Eq_psi}:
\begin{equation}
\psi_i = \sum_{b}\eta_{b}\phi_b.
\end{equation}
With this information, a full characterization of the model for a specific material can be performed, followed by subsequent simulation of ARPES matrix elements. This includes density of states, bandstructure, 3D Fermi surface, orbital projections of the eigenstates, as well as the expectation values of various operators of interest, such as for example, $\left< \bf L\cdot S\right>$ as in Fig. \ref{Fig:Fig2} for tetragonal FeSe. By defining an $N\times N$ Hermitian matrix, the expectation value of any observable operator can be computed in this way.

\subsection{Computation of Matrix Elements}
\begin{figure}[t!]
\includegraphics[width=\columnwidth]{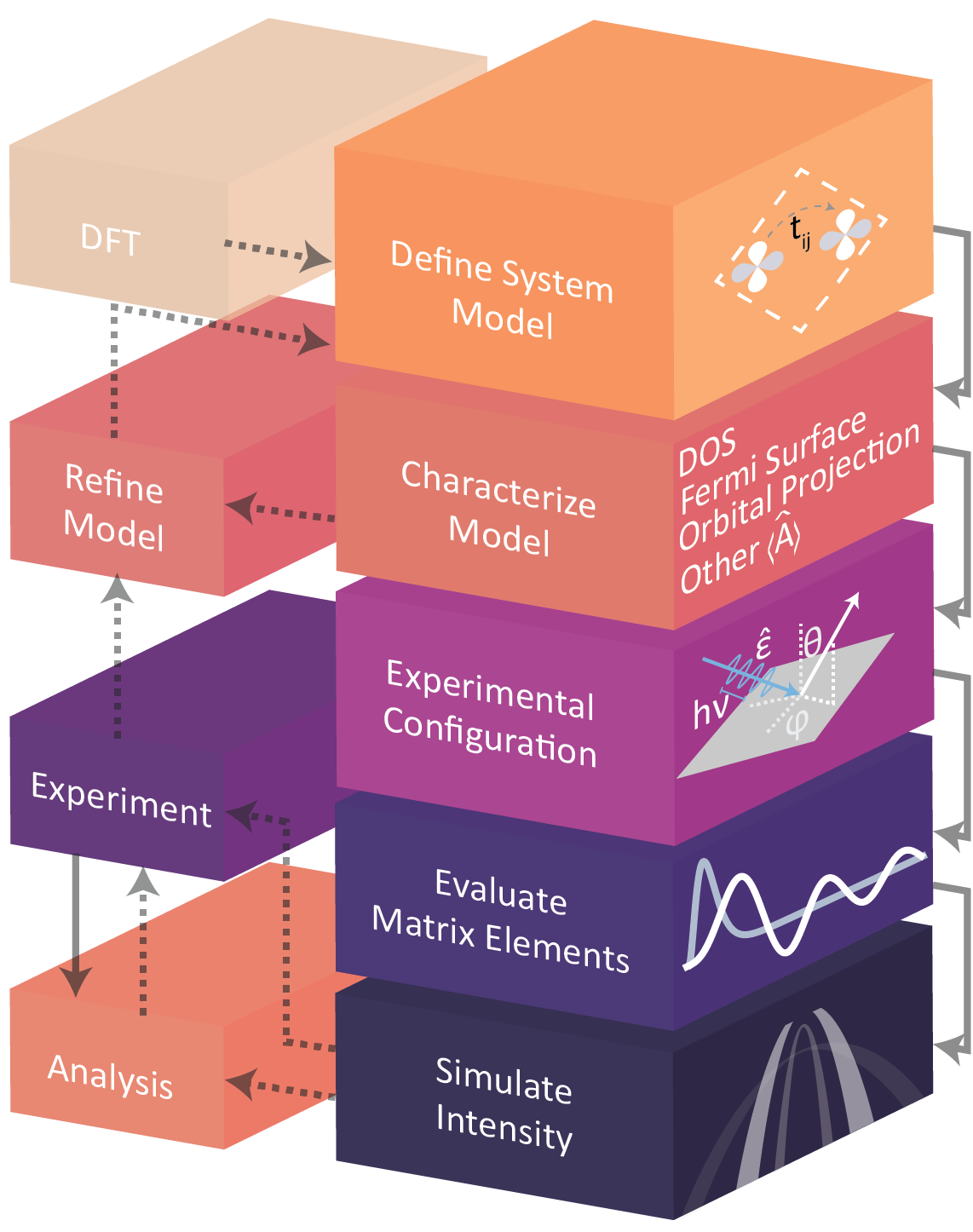}
\caption{$chinook$ workflow. Essential steps are denoted by solid dark lines, with dashed arrows indicating optional iterative methods. Informed by experiment, DFT, or literature, the user defines a model Hamiltonian, including lattice geometry, orbital basis, and kinetic terms. This model can be characterized through consideration of the (orbitally-projected) bandstructure, total and projected density of states (DOS), Fermi surface, and the expectation value of other relevant operators. The experimental configuration is then defined, and the matrix element integrals can be computed. The resulting intensity is plotted, and compared against experiment. The model can be further refined, or the results exported for additional analysis.}
\label{Fig:Fig3}
\end{figure}

The workflow of $chinook$ is sketched in Fig. \ref{Fig:Fig3}. Once a satisfactory material model is established, one can proceed to the simulation of ARPES intensity maps. A suitable region of interest in momentum and energy space must be defined, and the eigenvalue problem is then solved over this domain.

We model the matrix elements of the dipole operator as:
\begin{equation}\label{eq:eqM}
M_{\alpha}(\vec{k},\omega) \propto  \matrixel{e^{i\vec{k}\cdot\vec{r}}}{\uv\epsilon\cdot\vec{r}}{\psi_i^{\alpha}},
\end{equation}
where we have made use of the commutation relations to express the dipole operator in the position representation.

In explicit evaluation of the ARPES matrix element, we expand both the initial and final states as prescribed by Eqs. \ref{Eq:Rayleigh} and \ref{Eq:Eq_psi}, which allows us to express Eq. \ref{eq:eqM} as:

\begin{equation}\label{eq:eq_matel}
\begin{split}
M_{\alpha}(\vec{k},\omega) &\propto  \matrixel{e^{i\vec{k}\cdot\vec{r}}}{\uv\epsilon\cdot\vec{r}}{\psi_i^{\alpha}}\\
=\sum_{b}& c_{\alpha}^b(\vec{k},\omega)\eta_{b} \int d^3r e^{i\vec{k}\cdot\vec{r}-z_{b}/2\xi}\uv{\epsilon}\cdot\vec{r}R_{b}(r)Y_{l_b}^{m_b}(\Omega)\\
=\sum_{b,l'}& c_{\alpha}^b\eta_{b} Y_{l'}^{m'}(\Omega _k) (i)^{l'}\int dr j_{l'}(kr)r^3 R_{b}(r)\\ &\times \sum_{\mu} \epsilon_{\mu}\int d\Omega Y_{l'}^{m'}(\Omega)Y_1^{\mu}(\Omega)Y_{l_{b}}^{m_{b}}(\Omega).
\end{split}
\end{equation}
This sum over integrals can be expressed in compact form as:
\begin{equation}\label{eq:eq_matel_simp}
M_{\alpha}(\vec{k},\omega) \propto \sum_{\mu,b,l',m}\epsilon_{\mu} c_{\alpha}^b(\vec{k},\omega)\eta_{b} Y_{l'}^{m}(\Omega_k) B_{b}^{l'}(k) G_{l',m}^{b,\mu},
\end{equation}

where $c_{\alpha}^b(\vec{k},\omega)\equiv\left<\phi_b|\psi_i^{\alpha} \right>$, and $\epsilon_{\mu}$ the components of the polarization vector. In the third line, we have absorbed an extinction factor $e^{-z_{b}/2\xi}$ into $\eta_{b}$, where $\xi$ represents the mean-free path of a photoemitted electron, and $z_{b}$ the spatial extent of the basis orbital below the surface \cite{Seah,Jason}.The radial and angular integrals are contained in the:
\begin{equation}\label{eq:radint}
B_{b}^{l'}(k)= (i)^{l'}\int dr j_{l'}(kr)r^3 R_{b}(r),
\end{equation}
and :
\begin{equation}
G_{l',m'}^{b,\mu} = \int d\Omega Y_{l'}^{m'}(\Omega)Y_1^{\mu}(\Omega)Y_{l_{b}}^{m_{b}}(\Omega),
\end{equation}
terms respectively. Here, $G_{l',m'}^{b,\mu}$ is equivalent to a small subset of Gaunt coefficients, allowing for efficient and exact evaluation of this term.
Meanwhile, the radial integrals can not necessarily be expressed in an analytical form and must be computed numerically, as the radial wavefunction is loosely constrained in most tight-binding models \cite{Molodtsov}. Whether a hydrogenic, Slater, or more complex object should be employed to describe the radial wavefunction is left to the discretion of the user, as the best choice is somewhat dependent on the nature of the material and states of interest. The user is given the opportunity to select from a variety of initial state wavefunctions in addition to importing their own functions or radial integrals at the start of the calculation. This could be the Wannier function grid as generated by for example Wannier90 \cite{W90}. 

We note that elsewhere it is common to take advantage of the plane-wave final state to recast the matrix element as a polarization modulated Fourier transform of the initial state \cite{Fourier,moser}. Specifically,
\begin{equation}\label{eq:eqMFT}
M_{FT} \propto \matrixel{ e^{i\vec{k}\cdot\vec{r}}}{\uv{\nabla}\cdot\uv{\epsilon}}{\psi_i} = i\vec{k}\cdot\uv{\epsilon}\braket{e^{i\vec{k}\cdot\vec{r}}}{\psi_i}.
\end{equation}
$M_{FT}$ can be expanded as in Eq. \ref{eq:eq_matel_simp}, establishing some formal equivalence to Eq. \ref{eq:eqM}. However, one commonly observes qualitative deviation from experiment within this description, due to the form of the radial integrals $B_{b}^{l'}$ introduced above. In the Fourier representation, these are written as:
\begin{equation}\label{eq:fourrad}
B_{b}^{l'}(k) = (i)^l\int dr j_l(kr)r^2R_b(r).
\end{equation}
One can contrast this with Eq. \ref{eq:radint} which we employ in $chinook$.  It is made explicit in Eq. \ref{eq:fourrad} that the Fourier representation of the dipole operator imposes radial integrals which are independent of final state angular momentum. The implications for multi-orbital systems, and for those where $l>0$, are important as final state interference becomes relevant. This is ultimately why for a plane-wave final state, the position, rather than momentum representation of the dipole operator yields a better description of experiment. We emphasize that although the limited constraints of tight-binding imply that the integrals $B_{b}^{l'}$ are to some extent parameters of the calculation, support for distinct final state angular momentum cross-sections is essential to the success of the position representation used here. Furthermore, it offers a natural extension of our framework to scattering-final states, wherein the commutation relations required to establish Eq. \ref{eq:eqM} are more rigourously justified. Further discussion of these approximations can be found in the Supplementary Materials.

\begin{figure*}[t]
\includegraphics[width=\textwidth]{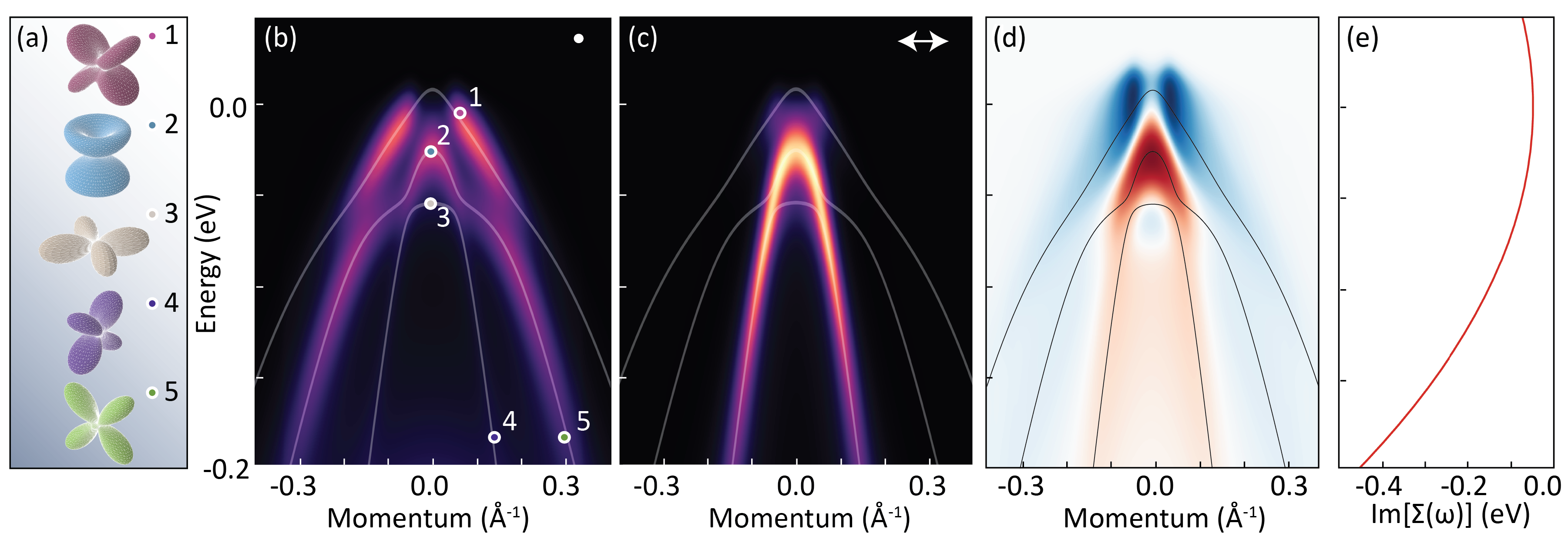}
\caption{Calculated ARPES spectra for FeSe. Performed with $h\nu = 37$ eV, along $\Gamma M$ direction. The balance of SOC and the crystal-field can be observed with the orbital projections plotted in (c) for several points along the dispersion, as indicated by the cursors in (b). In (b), (c), polarization vectors are indicated by arrows, corresponding to s-polarized $\uv{\epsilon} = [0,1,0]$ in (b) and p-polarized $\uv{\epsilon} = \sqrt{\frac{1}{2}}[-1,0,1]$ light in (c). The sample is aligned with the Fe-Fe bond direction oriented along the Cartesian basis.  In (d), calculated $A_z$ as from circularly-polarized spin-ARPES (CPS-ARPES) provides a more direct perspective on SOC, with an explicit connection to the $\left<\vec{L}\cdot\vec{S}\right>$, projected along the $\uv{z}$ axis. Finite linewidth of the spectra in (b,c,d) reflect the convolution of experimental resolution $\Delta E = 10$ meV, $\Delta k = 0.01$ \AA$^{-1}$ and Im[$\Sigma(\omega)$] plotted in (e).}
\label{Fig:Fig4}
\end{figure*}
Returning to the calculations executed in $chinook$, the central object of importance is the coherent matrix element factor:
\begin{equation}
M_{\alpha}(\vec{k}) = \begin{bmatrix} M_{\mu=-1}^{\downarrow} & M_{\mu=0}^{\downarrow} & M_{\mu=1}^{\downarrow} \\ M_{\mu=-1}^{\uparrow} & M_{\mu=0}^{\uparrow} & M_{\mu=1}^{\uparrow}  \end{bmatrix}_{\alpha}(\vec{k}).
\end{equation}
Evaluation of this object proceeds following Eq. \ref{eq:eq_matel}. Each column corresponds to the projection of the polarization vector in the basis of spherical harmonics, (i.e. $\mu\equiv\Delta m=\pm1,0$), and the rows indicate the spinor projections. By retaining the matrix element in this coherent form, the ARPES intensity for arbitrary polarization and spin projection can be recalculated at run-time with minimal computational overhead. For each band and k-point in the region of interest, a spectral function as defined in Eq. \ref{eq:spectral} is added to the total intensity map, with its amplitude multiplied by: 
\begin{equation}
|M_{\alpha}|^2 = |\sum_{\mu} \epsilon_\mu M^{\downarrow}_{\alpha,\mu}(k)|^2 + |\sum_{\mu} \epsilon_\mu M^{\uparrow}_{\alpha,\mu}(k)|^2.
\end{equation}
 The photoemission intensity is then computed as described in Eq. \ref{eq:eq_fgr}, with $|\bra{\psi_f}\uv{\Delta}\ket{\psi_i}|^2\rightarrow|M_{\alpha}|^2$.  Spin projection, polarization, resolution, temperature and self-energy can all be updated with little overhead at run-time. 

With ARPES intensity maps then calculated for different experimental configurations, these results can be exported for further analysis, or combined to define quantities such as spin-polarization and circular/linear dichroism. In this sense, the output of the standard $chinook$ calculation is a three-dimensional array of intensity in coordinates of momentum and energy which can be explored and analyzed in the same way as an experimental ARPES measurement. 

\section{Applications}

\subsection{Bulk Electronic Structure and Orbital Texture}
Returning to the motivating case of the Fe-based superconductor FeSe, we can implement the model characterized above in Fig. \ref{Fig:Fig2} and compare the simulated ARPES intensity against the low energy region of Fig. \ref{Fig:Fig1}. As with the experiments, the calculations were done at $h\nu=37$ eV and  $T=120$ K. A Fermi-liquid type self-energy has been applied to the spectral features, resulting in an energy-dependent broadening of the photoemission linewidth. In the present case, the tight-binding model has already been renormalized to match the experimental spectra, such that the  dispersion is more appropriately defined as $\epsilon_k' = \epsilon_k^0 -\Sigma'(k,\omega)$. Consequently, the self-energy used in the ARPES simulation is purely imaginary, $\Sigma(\vec{k},\omega) = i\Sigma"(\omega) = -i(0.005 + 1.0 \omega^2)$, which is plotted in Fig. \ref{Fig:Fig4}e. As ARPES matrix-elements can confound the evaluation of the spectral function and correlation effects in experimental data \cite{Borisenko_2003,sublattice,sigma_se}, the ability to model both components in the same environment can facilitate the disentanglement of these two objects of interest.

The simulation in Fig. \ref{Fig:Fig4} captures the relative intensity ratio between the three hole bands, with the heaviest (largest effective mass) band visible only through the SOC-induced hybridization gaps near $E_{B}=50$ meV. This latter state, composed primarily of $d_{xy}$ orbitals, has vanishing photoemission intensity along the normal emission ($\lim_{k_{\|}\rightarrow0}$) direction due to the selection rules associated with its definition in terms of spherical harmonics $Y_{2}^{\pm2}$: all possible final states have a node along the normal emission direction. While conventional interpretation of the remaining states assumes $d_{xz/yz}$-like wavefunctions, SOC allows for finite intensity from both states, as observed both experimentally and in the simulation near $k_{\|}=0$ \AA$^{-1}$. These inferences regarding the orbital structure of the initial states are supported by projection of the tight-binding eigenstates onto the basis of spherical harmonics, as done at select k-points in Fig. \ref{Fig:Fig4}a using built-in diagnostic tools from $chinook$. 

A more direct measure of the influence of SOC can be achieved through combining circularly polarized light with spin resolution to gain explicit access to both spin and orbital degrees of freedom. One can define the polarization asymmetry as:
\begin{equation}
A_z = \frac{\sqrt{I_{+}^{\downarrow}I_{-}^{\uparrow}} - \sqrt{I_{+}^{\uparrow}I_{-}^{\downarrow}}}{\sqrt{I_{+}^{\downarrow}I_{-}^{\uparrow}} + \sqrt{I_{+}^{\uparrow}I_{-}^{\downarrow}}},
\end{equation}
where subscripts indicate the helicity of light polarization, and superscripts the spin-projection of the photoelectrons. This quantity is closely related to the projection of $\left<\vec{L}\cdot\vec{S}\right>$ along the quantization axis of the experiment, allowing for a connection between Fig. \ref{Fig:Fig2}c and Fig. \ref{Fig:Fig4}d to be made. This technique has been applied to both ruthenates \cite{Mizokawa,veenstra} and Fe-based superconductors \cite{Day}, and utilizes the dipole selection rules encoded within the matrix element factor to provide the most direct measure of spin-orbital entanglement in solid state.

\subsection{Supercell Impurity Model}

Consideration of a supercell model illustrates the information encoded in the ARPES matrix element beyond orbital symmetry alone. Regarding the electronic structure of periodic systems, one can choose an arbitrarily large unit cell in exchange for a reduced Brillouin zone and additional backfolded bands. By contrast, impurities or other symmetry-breaking potentials (SBP) explicitly require such an expanded unit cell. While one can numerically perform an unfolding of these bands in an attempt to recover the spectrum within the extended Brillouin zone \cite{WKu_unfold}, such an unfolding is carried out naturally in the photoemission experiment. 

In the absence of the SBP, Bloch's theorem would impose that the original band becomes a symmetric superposition over the neighbouring lattice sites. The additional bands, which must be orthogonal to the original state will destructively interfere in evaluation of the ARPES matrix element, preventing observation of many of the folded states. Ultimately, an SBP can mix these states; when the SBP is an essential feature of the potential landscape, as in graphene and the Fe-based superconductors \cite{Chiang,WKu_12}, the folded bands can be observed with strong intensity over a range of momenta. When the SBP is weak or disordered, intensity from these folded bands vanishes away from the avoided crossings.
\begin{figure}[t!]
\includegraphics[width=\columnwidth]{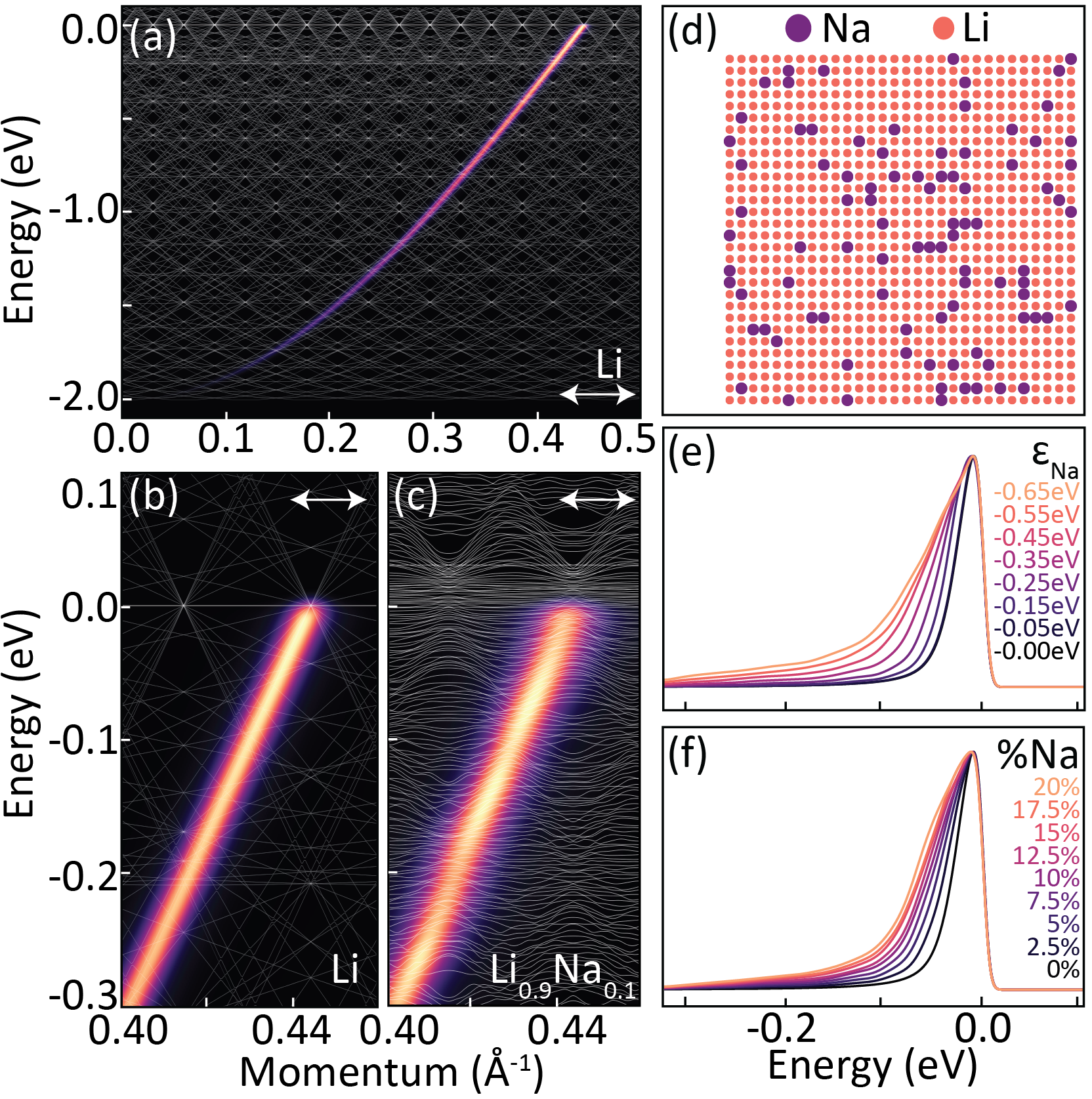}
\caption{Na impurity-substituted Li supercell. In (a-c), we plot the ARPES intensity at 21.2 eV with s-polarized (along momentum axis) light over the $\Gamma M$ direction of the extended Brillouin zone. The spectra have been averaged over 80 random configurations of a 30$\times$30 square lattice (as for example in (d)). Panels (a), (b) represent the pristine lattice of Li: the dispersion follows precisely that of the 1-Li unit cell, with all other states destructively interfering to produce zero intensity. Representative tight-binding bandstructures are plotted over the spectra. In (c) however, 90 Na atoms ($\epsilon_{Na} = -0.35$ eV) have been substituted for Li. The effect of impurity potential $\epsilon_{Na}$ is demonstrated for the series of EDCs at the Fermi momentum $k_F$, at a fixed concentration of 10$\%$ Na in (e). Similarly, fixing $\epsilon_{Na}=-0.35$ eV, the same is done for different concentrations in (f).}
\label{Fig:Fig5}
\end{figure}
To demonstrate these effects, we consider the artificial example of a square lattice of Li 2$s$ orbitals, into which we substitute some number of Na 3$s$ orbitals. Allowing for nearest neighbour hopping alone, and imposing a $\epsilon_{Na}$ = -0.35 eV impurity potential for the Na sites, we simulate the effect of local defects in this lattice and the resulting ARPES spectra. Kinetic terms and onsite potentials have been adapted from the phenomenological rules set out in Ref. \onlinecite{Harrison}. In an attempt to consider the impurity problem realistically, we populate a 30$\times$30 supercell of Li with various concentrations of randomly distributed Na impurities.  For each particular distribution, the density of states is integrated to fix the Fermi-level at half-filling, consistent with electron counting. We then compute the photoemission intensity at $h\nu= 21.2$ eV over the $\Gamma M$ direction of the extended Brillouin zone. For clarity, we assume a constant intrinsic linewidth of 10 meV along the entire dispersion. Energy and momentum resolution are set to 10 meV and 0.005 \AA$^{-1}$. The results, plotted in Fig. \ref{Fig:Fig5}c have been averaged over 80 such configurations, corresponding to a nominal doping of Li$_{0.9}$Na$_{0.1}$. This can be compared against the pure Li-supercell in Fig. \ref{Fig:Fig5}b. The full spectrum of the latter is displayed in Fig. \ref{Fig:Fig5}a. 

In each case, we compute the photoemission intensity from all states. However intensity from all folded bands is vanishing in the absense of the SBP; the full bandstructure is plotted in white over the spectra to demonstrate the large suppression of photoemission intensity. At the bottom of the band, the dipole selection rules suppress photoemission intensity from even the main band. In the disorder-averaged supercell, a substantial broadening of the spectral lineshape is observed \cite{maestro_defect,ast}.  As indicated by the overlain bandstructures of Fig. \ref{Fig:Fig5}a-c, the impurity potential introduces a high density of avoided crossings, where the eigenvector supports finite photoemission intensity. In this sense, the broadening can be associated with the relative phases within the tight-binding eigenvector to which the ARPES matrix element is sensitive. 

One can demonstrate that the linewidth broadening is dependent on both concentration and strength of impurities. In Fig. \ref{Fig:Fig5}e,f we plot energy distribution curves (EDCs) at $k_F=0.44$ \AA$^{-1}$ for fixed concentration (Li$_{0.9}$Na$_{0.1}$) with variable attractive (negative) $\epsilon_{Na}$, and fixed $\epsilon_{Na}=-0.35$ eV with variable concentration. Each spectrum has been averaged over 80 similar configurations, and normalized to its peak intensity. The linewidth is observed to increase monotonically with both concentration and impurity potential, indicating the similar role these degrees of freedom play in modifying the spectra of disordered systems. Despite this lineshape broadening, the low-energy dispersion is resilient against a high level of disorder, as illustrated by Fig. \ref{Fig:Fig5}c. By applying an out-of-plane polarization sensitive to states near the bottom of the band, we also confirm an increase of the bandwidth for this attractive impurity potential, which grows quadratically with impurity potential for the modest $\frac{|\epsilon_{Na}|}{W}<0.15$ considered here: at 10$\%$ Na and $\epsilon_{Na}=-0.35$ eV, the band bottom is extended 30 meV. Such detailed study of the impurity-substituted ARPES spectra is not possible without consideration of the matrix elements, which allow for a straightforward disentanglement of the supercell bandstructure and an opportunity to achieve meaningful insights from disordered materials.

\subsection{Surface vs Bulk, and Emergent $k_z$ Dispersion}

\begin{figure}[t!]
\includegraphics[width=\columnwidth]{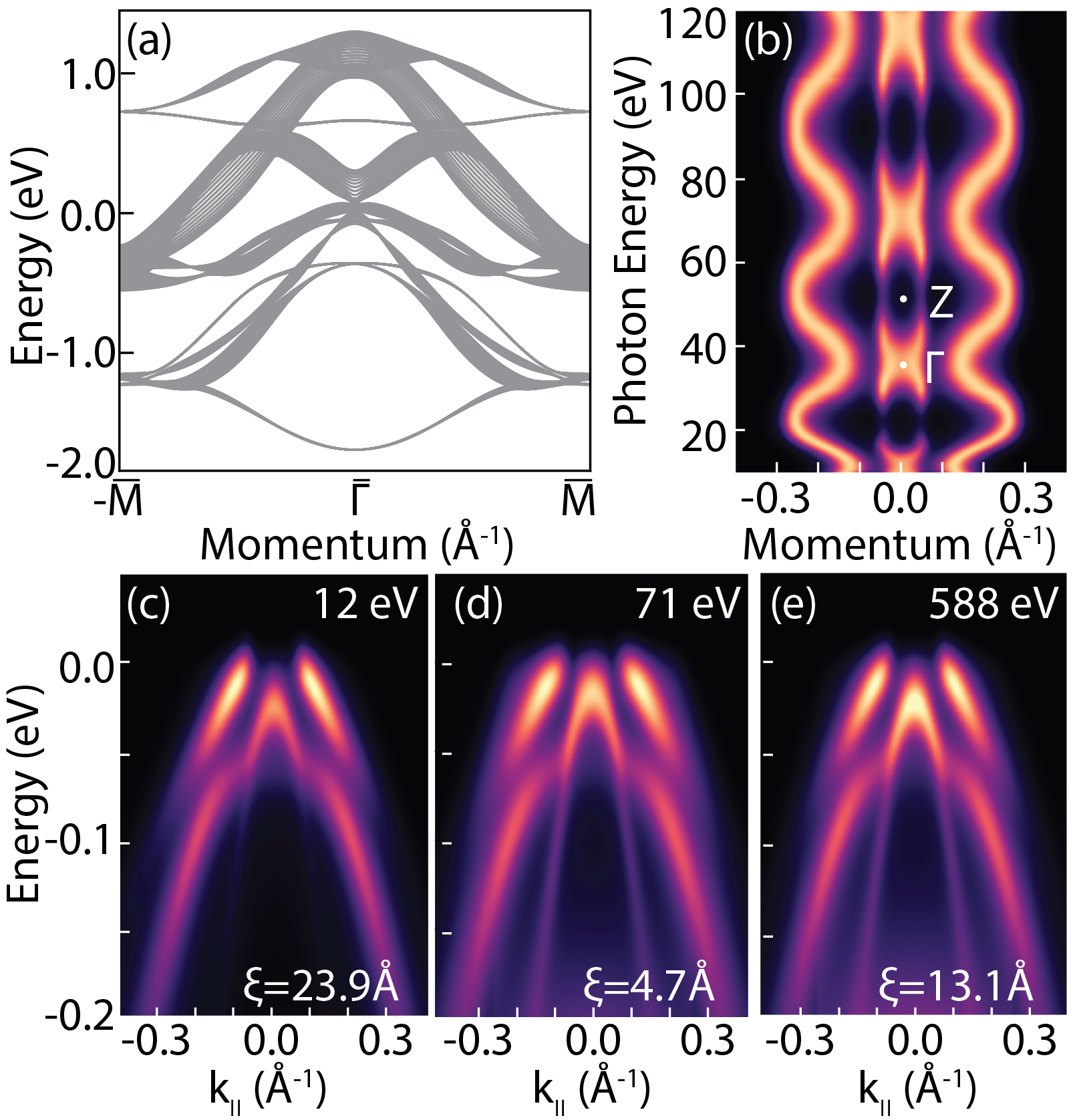}
\caption{Surface-projected photon energy dependence in FeSe. In (a), we build a 20 layer slab tight-binding model from the bulk model in Fig. \ref{Fig:Fig2}. We calculate the ARPES intensity along the $\Gamma M$ direction, at several photon energies; the results are summarized by a cut at constant binding energy of $E_B=25$ meV in panel (b). Spectra at each photon energy have been renormalized to their maximum intensity. In panels (c-e), spectra at select photon energies are plotted, chosen to correspond to the same $k_z=0$ \AA$^{-1}$ point in successive Brillouin zones to enable direct comparison. Photon energy dependence of the measured linewidth is observed due to effective $k_z$ integration.}
\label{Fig:Fig10}
\end{figure}

To this point we have considered the bulk-electronic structure, but it is important to appreciate the surface-sensitivity of the ARPES experiment: the high scattering cross-section in the ultraviolet regime results in penetration depths of the order of 5-10 \AA$ $ \cite{Seah}. This corresponds to the top few unit cells of the lattice, depending on experimental details. In many cases, the surface introduces modest corrections to the local electronic structure, facilitating a direct connection between the measured photoemission intensity and the bulk electronic structure \cite{DamascelliSRO}. In others, details of the surface preparation result in reconstructions of the ARPES spectra which deviate profoundly from the bulk electronic structure \cite{DamascelliSRO, Hossain, BorisenkoYBCO, Veenstra2013}. 

This surface sensitivity becomes rather important in the context of three- or even quasi-two- dimensional materials, where the photoelectron escape depth and $k_z$ information are intimately connected. For intermediate energies in the ultraviolet regime, where the penetration depth is of the order of 5 \AA, the $\Delta k_z$ required by the uncertainty principle becomes comparable to the size of the Brillouin zone. In the presence of finite $k_z$ dispersion, this can result in anomalously broad linewidths, as the spectrum effectively integrates over the third dimension of momentum space. This is visualized well in Fig. \ref{Fig:Fig10}, where we have projected our FeSe model onto a 20-layer slab model along the (001) direction. While the slab bandstructure is by construction independent of $k_z$, signatures closely related to the bulk $k_z$ dispersion are observed in photon energy dependent matrix element calculations, as seen in Fig. \ref{Fig:Fig10}b. We estimate the attenuation factor $e^{-\xi/2z_i}$ of the escaping photoelectrons using the universal escape depth curve from Ref. \onlinecite{Seah}. The $k_z$ value probed is calculated using an inner potential of $V_0 = 12.2$ eV \cite{watson}. While at both low and high photon energies the penetration depth is sufficiently large that that $\Delta k_z$ should be less than 0.05 \AA$^{-1}$, at $h\nu = 71$ eV $\Delta k_z = 0.11$ \AA$^{-1}$, and linewidth broadening is observed as a result (note here $\pi/c = 0.57$ \AA$^{-1}$). Conversely, for larger $\xi$ values, $\Delta k_z$ becomes negligible, and something akin to the bulk electronic structure is recovered. Note that $\Delta k_z$ is not explicitly included in these calculations, but emerges naturally from the combination of slab geometry with variable penetration depth. From these results, it becomes evident that the surface sensitivity can complicate successful estimation of the bulk electronic structure. This emphasizes the need for proper characterization of the $k_z$ dispersion, accessible via photon-energy dependent measurements, as in Fig. \ref{Fig:Fig10}b. 

\begin{figure}[t!]
\includegraphics[width=\columnwidth]{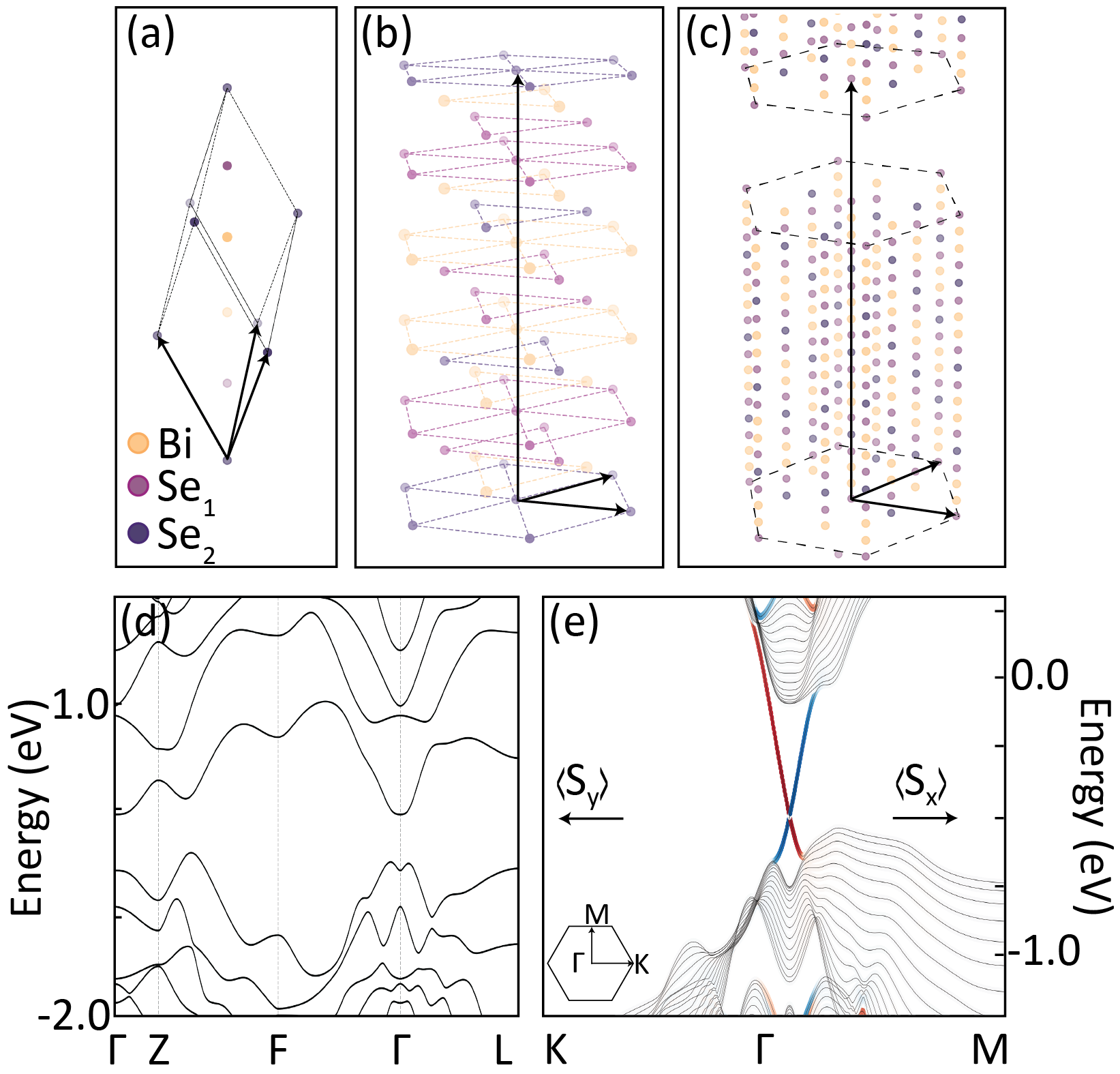}
\caption{Topological surface states in Bi$_2$Se$_3$. The progression from the rhombohedral unit cell of the bulk lattice (a) to the (111)-surface projected (b) hexagonal unit cell, and finally a Se$_1$-terminated slab (c) is plotted in the top row. In the bottom row, the bulk (d) and an 11-QL slab (e) bandstructure are compared. The colourscale of panel (e) reflects the expectation value of the surface-projected spin, directed orthogonal to the momentum axis ($\left<S_y\right>$ along $k_x$, and $\left<S_x\right>$ along $k_y$).}
\label{Fig:Fig6}
\end{figure}

\subsection{Photoelectron Interference, and Spin-ARPES}
Despite these challenges associated with surface sensitivity, the surface can also precipitate new states localized to the interface region which are not possible in bulk systems. Such is the case for example in the Shockley surface states observed along the (111)-termination of noble metals \cite{CuKKR,Tamai}, Fermi arcs on Weyl-semimetals \cite{Weyl}, and conductive surface states observed in topological insulators such as Bi$_2$Se$_3$ \cite{Hsieh}. 

To model the ARPES spectra from these surface states, an extended lattice basis is required, with the unit cell projected onto a slab-geometry.  Our implementation of the slab generation is inspired by the algorithm in Ref. \onlinecite{ceder}, allowing for nearly total automation of the slab Hamiltonian initialization. Given a surface Miller index, new lattice vectors can be defined which projects the new unit cell along the desired surface direction to the desired thickness. The bulk Hamiltonian can be propagated over this slab supercell. While we formally maintain periodic boundary conditions, rather than preserve the full translational symmetry of the bulk crystal, a vacuum buffer is defined, with a thickness sufficiently large to suppress hopping elements between neighbouring slab unit cells. The precise location of the crystal-vacuum interface is tuned by the user to achieve the desired surface termination of the crystal. In the case of Bi$_2$Se$_3$, this termination must occur between the van der Waals-bonded layers of two adjacent quintuple layers (QL) to preserve the topological surface states. The procedure is illustrated in Fig. \ref{Fig:Fig6}a-c.

\begin{figure}[t!]
\includegraphics[width=\columnwidth]{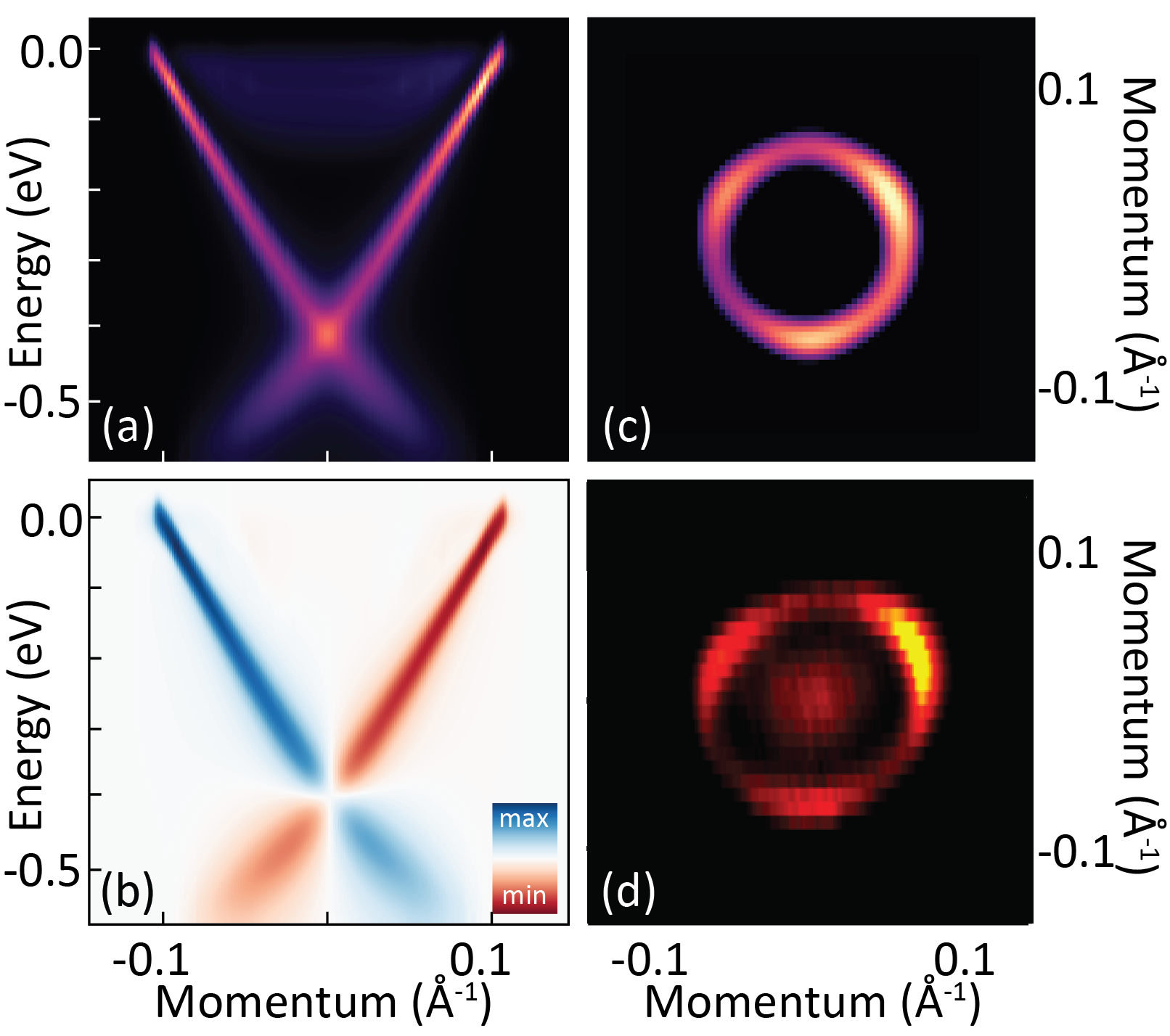}
\caption{ARPES spectra from topological surface states on Bi$_2$Se$_3$. In (a), simulated photoemission spectra along the $k_x$ direction around the surface Brillouin zone $\Gamma$ point, as observed with p-polarized light ($\uv{\epsilon} = \sqrt{\frac{1}{2}}[-1,0,1]$). Calculated spin-polarization, projected out of the plane of the page (c.f. Eq. \ref{Eq:Py}) for the same region is plotted in (b). Constant energy contours give evidence for bulk-hybridization and interlayer interference, as seen in both the simulation (c) and experiment (d). Data in panel (d) reproduced from \cite{Jason} with permission from the authors.}
\label{Fig:Fig7}
\end{figure}

Expansion of the basis set to a suitably large slab carries the caveat of a significant memory overhead, which can be to an extent mitigated in the calculation of ARPES intensity: as the finite penetration depth of the probe and photoemitted electrons limit the volume of the unit cell to which we are actually sensitive, the eigenvectors are truncated beyond a modest multiple of the mean free path, allowing for both efficient and high-fidelity surface-projected ARPES maps to be computed, as done for the 400-orbital basis used for the simulation in Fig. \ref{Fig:Fig7}a. As a result, it is the mean-free path more than the size of the basis which limits the ability to treat very large slab unit cells. As an example of this functionality, simulated and experimental ARPES intensity from Bi$_2$Se$_3$ are plotted in Fig. \ref{Fig:Fig7}. Many of the central tenets of a model strong TI have been confirmed in this material, such as the anticipated chiral spin texture, observed directly via spin-resolved ARPES \cite{Hsieh,gedik,Zhu_spin,Jozwiak}. Such spin-resolved experiments \cite{Osterwalder,Dil,Okuda} can also be simulated within the $chinook$ software, as shown in Fig. \ref{Fig:Fig7}b, where we present the simulated spin polarization:
\begin{equation}\label{Eq:Py}
P_y = \frac{I_{\uparrow_y} - I_{\downarrow_y}}{I_{\uparrow_y} + I_{\downarrow_y}}.
\end{equation}
This result is in agreement with experiment \cite{Hsieh}, and can be compared favourably with the surface-projected expectation value of the spin $\uv{S}$ operator, $e^{-\frac{|\uv{z}|}{\xi}}\uv{S}$, plotted in Fig \ref{Fig:Fig6}e. The bulk states, which lack any discernible spin-polarization (Fig. \ref{Fig:Fig6}d), vanish from the calculation of $P_y$ and so do not appear in Fig. \ref{Fig:Fig7}b.

While the topological surface states $\Psi_{TSS}$ are primarily composed of $p_z$ orbitals at the surface, a pronounced modulation of the photoemission intensity around the Dirac cone is observed as a consequence of the finite extension of $\Psi_{TSS}$ into the crystal bulk. Hybridization with bulk states, in addition to interlayer photoelectron interference can be understood as the progenitor of this modulation, as explored in depth in Ref. \onlinecite{Jason}. The interpretation of this angular intensity pattern in ARPES measurements presents an essential experimental verification and explanation of the limitations of applying a simple $\bf k\cdot p$ model to the description of real topological insulators such as Bi$_2$Se$_3$. While localized within a finite region near the vacuum interface, the full three-dimensionality of the surface state becomes apparent through consideration of this spectroscopic evidence. Convenient extension to a slab-geometry is then a critical functionality offered by the $chinook$ package.

\subsection{Variable Experimental Geometry}
In practice, ARPES experiments rotate either the sample normal or spectrometer in order to access a broad set of emission angles.  While some modern techniques such as photoemission electron microscopy (PEEM) \cite{kirschner_spin}, angle-resolved time-of-flight (ARTOF) \cite{gedik,gedik_floquet}, and deflector-based ARPES \cite{eli_deflector2,JozwiakTOF,deflector} apparatus avoid this complication, the assumption of a constant experimental geometry is not always possible. Furthermore, it is often advantageous to rotate the sample orientation in order to for example explore large regions of momentum space, or to achieve better momentum resolution available at higher emission angles \cite{scripta}. While the most direct complication is associated with variable photon polarization, in the case of S-ARPES, the relative orientation of the detector with respect to the sample is essential to interpret data correctly. 
\begin{figure*}[t!]
\includegraphics[width=\textwidth]{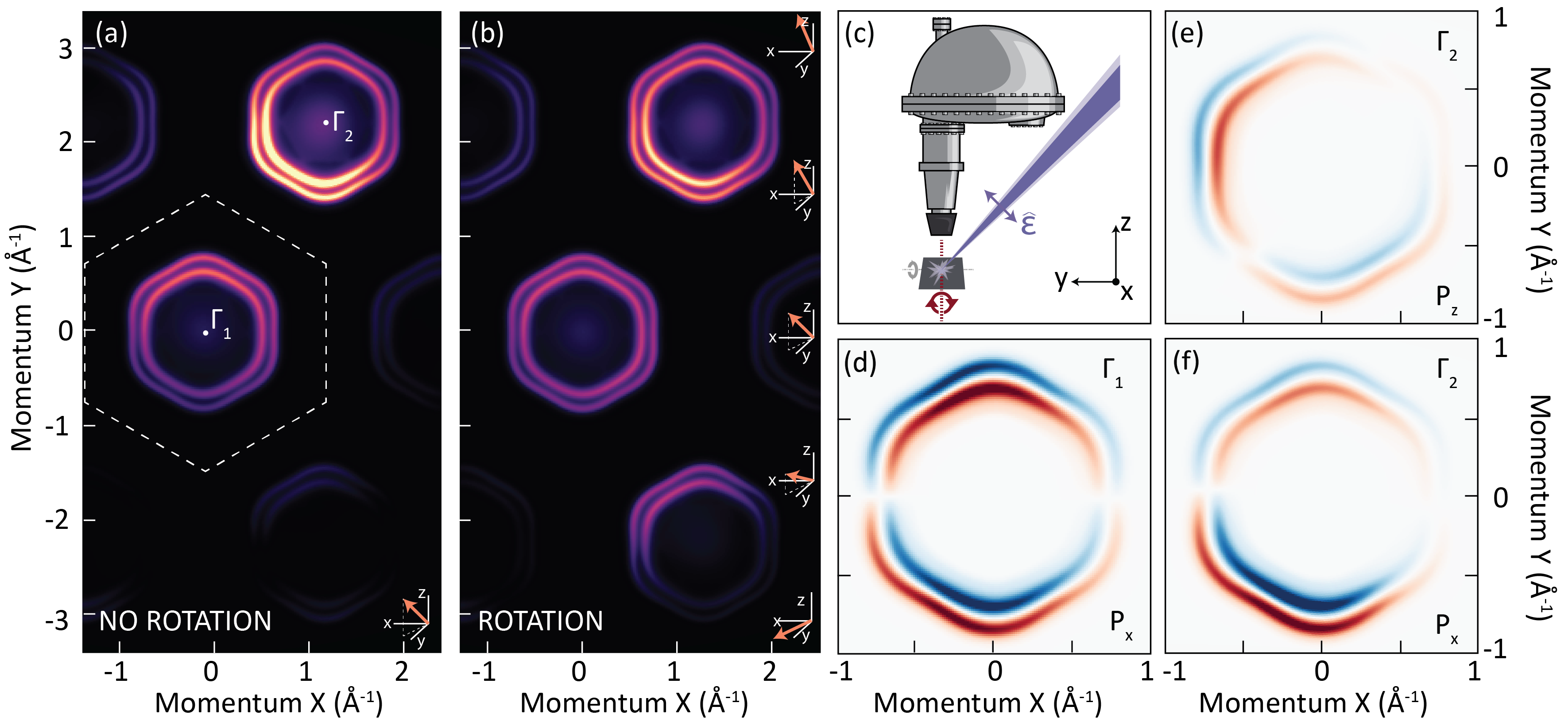}
\caption{Experimental geometry considerations. In panels (a) and (b), we plot the Fermi surface of PtCoO$_2$ over several neighbouring Brillouin zones, neglecting (a) and including (b) the rotation of the polarization vector associated with the sample orientation. The range of polarization vectors are illustrated by arrows in panel (b): for the horizontal ($k_x$) analyzer entrance slit, the polarization is fixed along lines of constant $k_y$. Moving from bottom to top, the sample rotates by $\approx$ 90$^o$, and the polarization goes from entirely in-plane, projected almost along the $\uv{y}$ direction, to almost entirely out of plane along the $\uv{z}$ direction. The geometry is drawn schematically in (c): the sample is rotated about the red ($\uv{x}$) axis to access the full domain of $k_y$, with the polarization as drawn: $\uv{\epsilon} = \sqrt{\frac{1}{2}}[0,1,1]$ in the laboratory frame. Finite $\epsilon_x$ at all angles results from a rotation of 7.2$^o$ about the grey axis in (c) to select the desired $k_x$ window. In (d-f) we compare the measured spin polarization in the first (d) and second Brillouin zones (e,f). Axis labels indicate distance from $\Gamma_{1,2}$. While a chiral Rashba spin texture is observed near normal emission, contamination between the different spin-channels is manifest as a substantial and artificial out-of plane spin projection in the second zone. $P_z$ is zero near $\Gamma_1$ and not shown here. All related colourscales are represented on the same scale to facilitate direct comparison.}
\label{Fig:Fig8}
\end{figure*}
To exemplify the practical considerations associated with such experiments wherein the geometry is variable, one may consider an exploration of Rashba-split spin-polarized surface states, as in for example PtCoO$_2$ \cite{sunko}. Using the model presented in Ref. \onlinecite{sunko}, in Fig. \ref{Fig:Fig8}, the Fermi surface is surveyed over several neighbouring Brillouin zones, accessed by rotation of the sample about the horizontal (i.e. $k_x$) axis. The ARPES intensity calculated with and without consideration of the rotated polarization vector can be compared in Fig. \ref{Fig:Fig8}a, b. While the intensity in the first Brillouin zone is fairly homogeneous in either case, the higher order zones reflect more substantial variation. These rotations complicate the extraction of orbital character from the photoemission intensity, requiring explicit consideration of the polarization rotation. 

In the context of spin-resolved measurements, the spin-projection is measured within the laboratory frame-of-reference, which remains fixed for all sample orientations. As the sample is rotated, contamination of orthogonal spin-channels is inevitable. While such effects are minimal near $\Gamma_1$, a significant out-of-plane spin-polarization arises near $\Gamma_2$, with $P_z$ over 36$\%$ of $P_x$, as demonstrated in Fig. \ref{Fig:Fig8}d-f. It is important to note that the intrinsic spin-polarization is confined entirely to the plane; this apparent out-of-plane polarization exists only in the coordinate frame of the laboratory apparatus. Accounting for the rotation of the measurement coordinate frame for the given experiment, one can then redistribute this information into the channels associated with the sample's intrinsic spin texture:
\begin{equation}
\vec{P_i} = \vec{P}_{exp} \cdot\vec{S(\theta)},
\end{equation}
where here $\vec{P}_{exp}$ is the measured spin polarization from Eq. \ref{Eq:Py} and $\vec{S(\theta)}$ the spin-projection axis measured at each emission angle. Although Fig. \ref{Fig:Fig8}b indicates higher photoemission intensity is available in the second Brillouin zone, Fig. \ref{Fig:Fig8}e,f illustrate the practical challenges associated with resolving the spin-texture near $\Gamma_2$. By affording the user with an ability to encode a realistic experimental configuration in the simulation, such effects can be accounted for in detail, circumventing a significant experimental limitation which may otherwise restrict more general application of the techniques detailed here.
\section{Conclusion}
We have presented here a simple and powerful numerical framework implemented in Python for the simulation and interpretation of ARPES spectra for a broad variety of materials of interest. Designed with this specific purpose in mind, the open-source structure of the $chinook$ software package is engineered to accommodate further extension beyond this application, as we have done recently for the study of resonant optical excitations in pump-probe spectroscopy experiments \cite{mxna}. Through the development of these tools, we hope to motivate and facilitate the consideration of the great depth of information encoded in the matrix element of angle-resolved photoemission spectroscopy towards a better understanding of these experiments and the electronic structure of the materials under consideration.
\section{Acknowledgements}
We happily acknowledge many helpful conversations with George A. Sawatzky and Elia Razzoli. This research was undertaken thanks in part to funding from the Max Planck-UBC-UTokyo Centre for Quantum Materials and the Canada First Research Excellence Fund, Quantum Materials and Future Technologies Program, in addition to the Killam, Alfred P. Sloan, and Natural Sciences and Engineering Research Council of Canada's (NSERC's) Steacie Memorial Fellowships (A.D.), the Alexander von Humboldt Fellowship (A.D.), the Canada Research Chairs Program (A.D.), NSERC, Canada Foundation for Innovation (CFI), British Columbia Knowledge Development Fund (BCKDF), and CIFAR Quantum Materials Program. 

\section{Author Contributions}
R.P.D. designed and wrote the code with support from B.Z. and I.S.E. The manuscript was written, and calculations performed by R.P.D.  All authors contributed to the interpretation of calculations and writing of the manuscript. I.S.E. and A.D. supervised the project.

\section{Additional Information}
\textbf{Competing Interests:} The Authors declare no Competing Financial or Non-Financial Interests.

\textbf{Data Availability:} Data presented here is available from R.P.D. on reasonable request.

\textbf{Code Availability:} The entire source code of $chinook$ is available at  \url{https://www.github.com/rpday/chinook}.

 \bibliographystyle{naturemag}
\bibliography{ubc_tbarpes}

\begin{thebibliography}{10}
\expandafter\ifx\csname url\endcsname\relax
  \def\url#1{\texttt{#1}}\fi
\expandafter\ifx\csname urlprefix\endcsname\relax\def\urlprefix{URL }\fi
\providecommand{\bibinfo}[2]{#2}
\providecommand{\eprint}[2][]{\url{#2}}

\bibitem{Mahan}
\bibinfo{author}{Mahan, G.~D.}
\newblock \bibinfo{title}{Theory of photoemission in simple metals}.
\newblock \emph{\bibinfo{journal}{Phys. Rev. B}} \textbf{\bibinfo{volume}{2}},
  \bibinfo{pages}{4334--4350} (\bibinfo{year}{1970}).

\bibitem{Hufner}
\bibinfo{author}{Hufner, S.}
\newblock \emph{\bibinfo{title}{Photoelectron Spectroscopy}}
  (\bibinfo{publisher}{Springer-Verlag}, \bibinfo{year}{1995}).

\bibitem{scripta}
\bibinfo{author}{Damascelli, A.}
\newblock \bibinfo{title}{Probing the electronic structure of complex systems
  by {ARPES}}.
\newblock \emph{\bibinfo{journal}{Physica Scripta}}
  \textbf{\bibinfo{volume}{2004}}, \bibinfo{pages}{61--74}
  (\bibinfo{year}{2004}).

\bibitem{Bansil}
\bibinfo{author}{Bansil, A.} \& \bibinfo{author}{Lindroos, M.}
\newblock \bibinfo{title}{Importance of matrix elements in the {ARPES} spectra
  of {BISCO}}.
\newblock \emph{\bibinfo{journal}{Phys. Rev. Lett.}}
  \textbf{\bibinfo{volume}{83}}, \bibinfo{pages}{5154--5157}
  (\bibinfo{year}{1999}).

\bibitem{shen}
\bibinfo{author}{Meevasana, W.} \emph{et~al.}
\newblock \bibinfo{title}{Hierarchy of multiple many-body interaction scales in
  high-temperature superconductors}.
\newblock \emph{\bibinfo{journal}{Phys. Rev. B}} \textbf{\bibinfo{volume}{75}},
  \bibinfo{pages}{174506} (\bibinfo{year}{2007}).

\bibitem{Gierz}
\bibinfo{author}{Gierz, I.}, \bibinfo{author}{Henk, J.},
  \bibinfo{author}{H\"ochst, H.}, \bibinfo{author}{Ast, C.~R.} \&
  \bibinfo{author}{Kern, K.}
\newblock \bibinfo{title}{Illuminating the dark corridor in graphene:
  Polarization dependence of angle-resolved photoemission spectroscopy on
  graphene}.
\newblock \emph{\bibinfo{journal}{Phys. Rev. B}} \textbf{\bibinfo{volume}{83}},
  \bibinfo{pages}{121408} (\bibinfo{year}{2011}).

\bibitem{Chiang}
\bibinfo{author}{Liu, Y.}, \bibinfo{author}{Bian, G.}, \bibinfo{author}{Miller,
  T.} \& \bibinfo{author}{Chiang, T.-C.}
\newblock \bibinfo{title}{Visualizing electronic chirality and {B}erry phases
  in graphene systems using photoemission with circularly polarized light}.
\newblock \emph{\bibinfo{journal}{Phys. Rev. Lett.}}
  \textbf{\bibinfo{volume}{107}}, \bibinfo{pages}{166803}
  (\bibinfo{year}{2011}).

\bibitem{Jason}
\bibinfo{author}{Zhu, Z.-H.} \emph{et~al.}
\newblock \bibinfo{title}{Layer-by-layer entangled spin-orbital texture of the
  topological surface state in {B}i${\mathrm{_{2}}}${S}e${\mathrm{_{3}}}$}.
\newblock \emph{\bibinfo{journal}{Phys. Rev. Lett.}}
  \textbf{\bibinfo{volume}{110}}, \bibinfo{pages}{216401}
  (\bibinfo{year}{2013}).

\bibitem{Gotlieb}
\bibinfo{author}{Gotlieb, K.} \emph{et~al.}
\newblock \bibinfo{title}{Symmetry rules shaping spin-orbital textures in
  surface states}.
\newblock \emph{\bibinfo{journal}{Phys. Rev. B}} \textbf{\bibinfo{volume}{95}},
  \bibinfo{pages}{245142} (\bibinfo{year}{2017}).

\bibitem{watson}
\bibinfo{author}{Watson, M.~D.} \emph{et~al.}
\newblock \bibinfo{title}{Emergence of the nematic electronic state in
  {F}e{S}e}.
\newblock \emph{\bibinfo{journal}{Phys. Rev. B}} \textbf{\bibinfo{volume}{91}},
  \bibinfo{pages}{155106} (\bibinfo{year}{2015}).

\bibitem{Day}
\bibinfo{author}{Day, R.~P.} \emph{et~al.}
\newblock \bibinfo{title}{Influence of spin-orbit coupling in iron-based
  superconductors}.
\newblock \emph{\bibinfo{journal}{Phys. Rev. Lett.}}
  \textbf{\bibinfo{volume}{121}}, \bibinfo{pages}{076401}
  (\bibinfo{year}{2018}).

\bibitem{dessau}
\bibinfo{author}{Cao, Y.} \emph{et~al.}
\newblock \bibinfo{title}{Mapping the orbital wavefunction of the surface
  states in three-dimensional topological insulators}.
\newblock \emph{\bibinfo{journal}{Nature Physics}}
  \textbf{\bibinfo{volume}{9}}, \bibinfo{pages}{499--504}
  (\bibinfo{year}{2013}).

\bibitem{veenstra}
\bibinfo{author}{Veenstra, C.~N.} \emph{et~al.}
\newblock \bibinfo{title}{Spin-orbital entanglement and the breakdown of
  singlets and triplets in {S}r${\mathrm{_{2}}}${R}u{O}${\mathrm{_{4}}}$
  revealed by spin- and angle-resolved photoemission spectroscopy}.
\newblock \emph{\bibinfo{journal}{Phys. Rev. Lett.}}
  \textbf{\bibinfo{volume}{112}}, \bibinfo{pages}{127002}
  (\bibinfo{year}{2014}).

\bibitem{CD_TI}
\bibinfo{author}{Xu, C.-Z.} \emph{et~al.}
\newblock \bibinfo{title}{Photoemission circular dichroism and spin
  polarization of the topological surface states in ultrathin
  {B}i${\mathrm{_{2}}}${T}e${\mathrm{_{3}}}$ films}.
\newblock \emph{\bibinfo{journal}{Phys. Rev. Lett.}}
  \textbf{\bibinfo{volume}{115}}, \bibinfo{pages}{016801}
  (\bibinfo{year}{2015}).

\bibitem{wathub}
\bibinfo{author}{Watson, M.~D.} \emph{et~al.}
\newblock \bibinfo{title}{Formation of {H}ubbard-like bands as a fingerprint of
  strong electron-electron interactions in {F}e{S}e}.
\newblock \emph{\bibinfo{journal}{Phys. Rev. B}} \textbf{\bibinfo{volume}{95}},
  \bibinfo{pages}{081106} (\bibinfo{year}{2017}).

\bibitem{Berend}
\bibinfo{author}{Zwartsenberg, B.} \emph{et~al.}
\newblock \bibinfo{title}{Spin-orbit controlled metal-insulator transition in
  {S}r${\mathrm{_{2}}}${I}r{O}${\mathrm{_{4}}}$}.
\newblock \emph{\bibinfo{journal}{Preprint available at arXiv:1903.00484}}
  (\bibinfo{year}{2019}).

\bibitem{Marchenko}
\bibinfo{author}{Marchenko, D.} \emph{et~al.}
\newblock \bibinfo{title}{Extremely flat band in bilayer graphene}.
\newblock \emph{\bibinfo{journal}{Science Advances}}
  \textbf{\bibinfo{volume}{4}}, \bibinfo{pages}{11} (\bibinfo{year}{2018}).

\bibitem{Grobman}
\bibinfo{author}{Grobman, W.~D.}
\newblock \bibinfo{title}{Angle-resolved photoemission from molecules in the
  independent-atomic-center approximation}.
\newblock \emph{\bibinfo{journal}{Physical Review B}}
  \textbf{\bibinfo{volume}{17}}, \bibinfo{pages}{4573--4585}
  (\bibinfo{year}{1978}).

\bibitem{moser}
\bibinfo{author}{Moser, S.}
\newblock \bibinfo{title}{An experimentalist's guide to the matrix element in
  angle resolved photoemission}.
\newblock \emph{\bibinfo{journal}{Journal of Electron Spectroscopy and Related
  Phenomena}} \textbf{\bibinfo{volume}{214}}, \bibinfo{pages}{29 -- 52}
  (\bibinfo{year}{2017}).

\bibitem{Kruger}
\bibinfo{author}{Kr{\"u}ger, P.}
\newblock \bibinfo{title}{Photoelectron diffraction from valence states of
  oriented molecules}.
\newblock \emph{\bibinfo{journal}{Journal of the Physical Society of Japan}}
  \textbf{\bibinfo{volume}{87}}, \bibinfo{pages}{061007}
  (\bibinfo{year}{2018}).

\bibitem{mulazzi}
\bibinfo{author}{Mulazzi, M.} \emph{et~al.}
\newblock \bibinfo{title}{Matrix element effects in angle-resolved valence band
  photoemission with polarized light from the {N}i(111) surface}.
\newblock \emph{\bibinfo{journal}{Phys. Rev. B}} \textbf{\bibinfo{volume}{74}},
  \bibinfo{pages}{035118} (\bibinfo{year}{2006}).

\bibitem{minar}
\bibinfo{author}{Min{\'a}r, J.}, \bibinfo{author}{Braun, J.},
  \bibinfo{author}{Mankovsky, S.} \& \bibinfo{author}{Ebert, H.}
\newblock \bibinfo{title}{Calculation of angle-resolved photoemission spectra
  within the one-step model of photoemission---recent developments}.
\newblock \emph{\bibinfo{journal}{Journal of Electron Spectroscopy and Related
  Phenomena}} \textbf{\bibinfo{volume}{184}}, \bibinfo{pages}{91 -- 99}
  (\bibinfo{year}{2011}).

\bibitem{CuKKR}
\bibinfo{author}{Winkelmann, A.} \emph{et~al.}
\newblock \bibinfo{title}{Analysis of the electronic structure of copper via
  two-dimensional photoelectron momentum distribution patterns}.
\newblock \emph{\bibinfo{journal}{New Journal of Physics}}
  \textbf{\bibinfo{volume}{14}}, \bibinfo{pages}{043009}
  (\bibinfo{year}{2012}).

\bibitem{EK_FeSC}
\bibinfo{author}{Eschrig, H.} \& \bibinfo{author}{Koepernik, K.}
\newblock \bibinfo{title}{Tight-binding models for the iron-based
  superconductors}.
\newblock \emph{\bibinfo{journal}{Phys. Rev. B}} \textbf{\bibinfo{volume}{80}},
  \bibinfo{pages}{104503} (\bibinfo{year}{2009}).

\bibitem{SK}
\bibinfo{author}{Slater, J.~C.} \& \bibinfo{author}{Koster, G.~F.}
\newblock \bibinfo{title}{Simplified {LCAO} method for the periodic potential
  problem}.
\newblock \emph{\bibinfo{journal}{Phys. Rev.}} \textbf{\bibinfo{volume}{94}},
  \bibinfo{pages}{1498--1524} (\bibinfo{year}{1954}).

\bibitem{vafek}
\bibinfo{author}{Cvetkovic, V.} \& \bibinfo{author}{Vafek, O.}
\newblock \bibinfo{title}{Space group symmetry, spin-orbit coupling, and the
  low-energy effective {H}amiltonian for iron-based superconductors}.
\newblock \emph{\bibinfo{journal}{Phys. Rev. B}} \textbf{\bibinfo{volume}{88}},
  \bibinfo{pages}{134510} (\bibinfo{year}{2013}).

\bibitem{Seah}
\bibinfo{author}{Seah, M.~P.} \& \bibinfo{author}{Dench, W.~A.}
\newblock \bibinfo{title}{Quantitative electron spectroscopy of surfaces: A
  standard data base for electron inelastic mean free paths in solids}.
\newblock \emph{\bibinfo{journal}{Surface and Interface Analysis}}
  \textbf{\bibinfo{volume}{1}}, \bibinfo{pages}{2--11} (\bibinfo{year}{1979}).

\bibitem{Molodtsov}
\bibinfo{author}{Molodtsov, S.~L.} \emph{et~al.}
\newblock \bibinfo{title}{Cooper minima in the photoemission spectra of
  solids}.
\newblock \emph{\bibinfo{journal}{Phys. Rev. Lett.}}
  \textbf{\bibinfo{volume}{85}}, \bibinfo{pages}{4184--4187}
  (\bibinfo{year}{2000}).

\bibitem{W90}
\bibinfo{author}{Mostofi, A.~A.} \emph{et~al.}
\newblock \bibinfo{title}{An updated version of {W}annier90: A tool for
  obtaining maximally-localised {W}annier functions}.
\newblock \emph{\bibinfo{journal}{Computer Physics Communications}}
  \textbf{\bibinfo{volume}{185}}, \bibinfo{pages}{2309 -- 2310}
  (\bibinfo{year}{2014}).

\bibitem{Fourier}
\bibinfo{author}{Puschnig, P.} \& \bibinfo{author}{L{\"u}ftner, D.}
\newblock \bibinfo{title}{Simulation of angle-resolved photoemission spectra by
  approximating the final state by a plane wave: From graphene to polycyclic
  aromatic hydrocarbon molecules}.
\newblock \emph{\bibinfo{journal}{Journal of Electron Spectroscopy and Related
  Phenomena}} \textbf{\bibinfo{volume}{200}}, \bibinfo{pages}{193 -- 208}
  (\bibinfo{year}{2015}).

\bibitem{Borisenko_2003}
\bibinfo{author}{Borisenko, S.~V.} \emph{et~al.}
\newblock \bibinfo{title}{Anomalous enhancement of the coupling to the magnetic
  resonance mode in underdoped {P}b-{B}i2212}.
\newblock \emph{\bibinfo{journal}{Phys. Rev. Lett.}}
  \textbf{\bibinfo{volume}{90}}, \bibinfo{pages}{207001}
  (\bibinfo{year}{2003}).

\bibitem{sublattice}
\bibinfo{author}{Jung, S.~W.} \emph{et~al.}
\newblock \bibinfo{title}{Sublattice interference as the origin of
  $\ensuremath{\sigma}$ band kinks in graphene}.
\newblock \emph{\bibinfo{journal}{Phys. Rev. Lett.}}
  \textbf{\bibinfo{volume}{116}}, \bibinfo{pages}{186802}
  (\bibinfo{year}{2016}).

\bibitem{sigma_se}
\bibinfo{author}{Mazzola, F.} \emph{et~al.}
\newblock \bibinfo{title}{Strong electron-phonon coupling in the
  $\ensuremath{\sigma}$ band of graphene}.
\newblock \emph{\bibinfo{journal}{Phys. Rev. B}} \textbf{\bibinfo{volume}{95}},
  \bibinfo{pages}{075430} (\bibinfo{year}{2017}).

\bibitem{Mizokawa}
\bibinfo{author}{Mizokawa, T.} \emph{et~al.}
\newblock \bibinfo{title}{{S}pin-orbit coupling in the {M}ott insulator
  {C}a$_2${R}u{O}$_4$}.
\newblock \emph{\bibinfo{journal}{Phys. Rev. Lett.}}
  \textbf{\bibinfo{volume}{87}}, \bibinfo{pages}{077202}
  (\bibinfo{year}{2001}).

\bibitem{WKu_unfold}
\bibinfo{author}{Ku, W.}, \bibinfo{author}{Berlijn, T.} \&
  \bibinfo{author}{Lee, C.-C.}
\newblock \bibinfo{title}{Unfolding first-principles band structures}.
\newblock \emph{\bibinfo{journal}{Phys. Rev. Lett.}}
  \textbf{\bibinfo{volume}{104}}, \bibinfo{pages}{216401}
  (\bibinfo{year}{2010}).

\bibitem{WKu_12}
\bibinfo{author}{Brouet, V.} \emph{et~al.}
\newblock \bibinfo{title}{Impact of the two {F}e unit cell on the electronic
  structure measured by {ARPES} in iron pnictides}.
\newblock \emph{\bibinfo{journal}{Phys. Rev. B}} \textbf{\bibinfo{volume}{86}},
  \bibinfo{pages}{075123} (\bibinfo{year}{2012}).

\bibitem{Harrison}
\bibinfo{author}{Harrison, W.~A.}
\newblock \emph{\bibinfo{title}{Electronic Structure and the Properties of
  Solids: The Physics of the Chemical Bond}} (\bibinfo{publisher}{Dover},
  \bibinfo{year}{1980}).

\bibitem{maestro_defect}
\bibinfo{author}{Kastl, C.} \emph{et~al.}
\newblock \bibinfo{title}{Effects of defects on band structure and excitons in
  {WS}${\mathrm{_{2}}}$ revealed by nanoscale photoemission spectroscopy}.
\newblock \emph{\bibinfo{journal}{ACS Nano}} \textbf{\bibinfo{volume}{13}},
  \bibinfo{pages}{1284--1291} (\bibinfo{year}{2019}).

\bibitem{ast}
\bibinfo{author}{Kot, P.}, \bibinfo{author}{Parnell, J.},
  \bibinfo{author}{Habibian, C., S.~Stra{\ss}er}, \bibinfo{author}{Ostrovsky,
  P.~M.} \& \bibinfo{author}{Ast, C.~R.}
\newblock \bibinfo{title}{Band dispersion of graphene with structural defects}.
\newblock \emph{\bibinfo{journal}{arxiv:1811.00087}}  (\bibinfo{year}{2018}).

\bibitem{DamascelliSRO}
\bibinfo{author}{Damascelli, A.} \emph{et~al.}
\newblock \bibinfo{title}{Fermi surface, surface states, and surface
  reconstruction in {S}r${\mathrm{_{2}}}${R}u{O}${\mathrm{_{4}}}$}.
\newblock \emph{\bibinfo{journal}{Phys. Rev. Lett.}}
  \textbf{\bibinfo{volume}{85}}, \bibinfo{pages}{5194--5197}
  (\bibinfo{year}{2000}).

\bibitem{Hossain}
\bibinfo{author}{Hossain, M.~A.} \emph{et~al.}
\newblock \bibinfo{title}{In situ doping control of the surface of
  high-temperature superconductors}.
\newblock \emph{\bibinfo{journal}{Nature Physics}}
  \textbf{\bibinfo{volume}{4}}, \bibinfo{pages}{527--531}
  (\bibinfo{year}{2008}).

\bibitem{BorisenkoYBCO}
\bibinfo{author}{Zabolotnyy, V.~B.} \emph{et~al.}
\newblock \bibinfo{title}{Momentum and temperature dependence of
  renormalization effects in the high-temperature superconductor
  {YB}a${\mathrm{_2}}${C}u${\mathrm{_3}}${O}${\mathrm{_{7-\delta}}}$.}
\newblock \emph{\bibinfo{journal}{Phys. Rev. B}} \textbf{\bibinfo{volume}{76}},
  \bibinfo{pages}{064519} (\bibinfo{year}{2007}).

\bibitem{Veenstra2013}
\bibinfo{author}{Veenstra, C.~N.} \emph{et~al.}
\newblock \bibinfo{title}{Determining the surface-to-bulk progression in the
  normal-state electronic structure of
  {S}r${\mathrm{_{2}}}${R}u{O}${\mathrm{_{4}}}$ by angle-resolved photoemission
  and density functional theory}.
\newblock \emph{\bibinfo{journal}{Phys. Rev. Lett.}}
  \textbf{\bibinfo{volume}{110}}, \bibinfo{pages}{097004}
  (\bibinfo{year}{2013}).

\bibitem{Tamai}
\bibinfo{author}{Tamai, A.} \emph{et~al.}
\newblock \bibinfo{title}{Spin-orbit splitting of the {S}hockley surface state
  on {C}u(111)}.
\newblock \emph{\bibinfo{journal}{Phys. Rev. B}} \textbf{\bibinfo{volume}{87}},
  \bibinfo{pages}{075113} (\bibinfo{year}{2013}).

\bibitem{Weyl}
\bibinfo{author}{Xu, S.-Y.} \emph{et~al.}
\newblock \bibinfo{title}{Discovery of a {W}eyl fermion semimetal and
  topological {F}ermi arcs}.
\newblock \emph{\bibinfo{journal}{Science}} \textbf{\bibinfo{volume}{349}},
  \bibinfo{pages}{613--617} (\bibinfo{year}{2015}).

\bibitem{Hsieh}
\bibinfo{author}{Hsieh, D.} \emph{et~al.}
\newblock \bibinfo{title}{A topological {D}irac insulator in a quantum spin
  {H}all phase}.
\newblock \emph{\bibinfo{journal}{Nature}} \textbf{\bibinfo{volume}{452}},
  \bibinfo{pages}{970--974} (\bibinfo{year}{2008}).

\bibitem{ceder}
\bibinfo{author}{Sun, W.} \& \bibinfo{author}{Ceder, G.}
\newblock \bibinfo{title}{Efficient creation and convergence of surface slabs}.
\newblock \emph{\bibinfo{journal}{Surface Science}}
  \textbf{\bibinfo{volume}{617}}, \bibinfo{pages}{53 -- 59}
  (\bibinfo{year}{2013}).

\bibitem{gedik}
\bibinfo{author}{Wang, Y.~H.} \emph{et~al.}
\newblock \bibinfo{title}{Observation of a warped helical spin texture in
  {B}i$_2${S}e$_3$ from circular dichroism angle-resolved photoemission
  spectroscopy}.
\newblock \emph{\bibinfo{journal}{Phys. Rev. Lett.}}
  \textbf{\bibinfo{volume}{107}}, \bibinfo{pages}{207602}
  (\bibinfo{year}{2011}).

\bibitem{Zhu_spin}
\bibinfo{author}{Zhu, Z.-H.} \emph{et~al.}
\newblock \bibinfo{title}{Photoelectron spin-polarization control in the
  topological insulator {B}i${\mathrm{_{2}}}${S}e${\mathrm{_{3}}}$}.
\newblock \emph{\bibinfo{journal}{Phys. Rev. Lett.}}
  \textbf{\bibinfo{volume}{112}}, \bibinfo{pages}{076802}
  (\bibinfo{year}{2014}).

\bibitem{Jozwiak}
\bibinfo{author}{Jozwiak, C.} \emph{et~al.}
\newblock \bibinfo{title}{Spin-polarized surface resonances accompanying
  topological surface state formation}.
\newblock \emph{\bibinfo{journal}{Nature Communications}}
  \textbf{\bibinfo{volume}{7}}, \bibinfo{pages}{13143} (\bibinfo{year}{2016}).

\bibitem{Osterwalder}
\bibinfo{author}{Osterwalder, J.}
\newblock \emph{\bibinfo{title}{Magnetism: A Synchrotron Radiation Approach}},
  vol. \bibinfo{volume}{697}, chap. \bibinfo{chapter}{Spin-Polarized
  Photoemission} (\bibinfo{publisher}{Springer}, \bibinfo{year}{2006}).

\bibitem{Dil}
\bibinfo{author}{Dil, J.~H.}
\newblock \bibinfo{title}{Spin and angle resolved photoemission on non-magnetic
  low-dimensional systems}.
\newblock \emph{\bibinfo{journal}{Journal of Physics: Condensed Matter}}
  \textbf{\bibinfo{volume}{21}}, \bibinfo{pages}{403001}
  (\bibinfo{year}{2009}).

\bibitem{Okuda}
\bibinfo{author}{Okuda, T.}
\newblock \bibinfo{title}{Recent trends in spin-resolved photoelectron
  spectroscopy}.
\newblock \emph{\bibinfo{journal}{Journal of Physics: Condensed Matter}}
  \textbf{\bibinfo{volume}{29}}, \bibinfo{pages}{483001}
  (\bibinfo{year}{2017}).

\bibitem{kirschner_spin}
\bibinfo{author}{Tusche, C.}, \bibinfo{author}{Krasyuk, A.} \&
  \bibinfo{author}{Kirschner, J.}
\newblock \bibinfo{title}{Spin resolved bandstructure imaging with a high
  resolution momentum microscope}.
\newblock \emph{\bibinfo{journal}{Ultramicroscopy}}
  \textbf{\bibinfo{volume}{159}}, \bibinfo{pages}{520 -- 529}
  (\bibinfo{year}{2015}).

\bibitem{gedik_floquet}
\bibinfo{author}{Wang, Y.~H.}, \bibinfo{author}{Steinberg, H.},
  \bibinfo{author}{Jarillo-Herrero, P.} \& \bibinfo{author}{Gedik, N.}
\newblock \bibinfo{title}{Observation of {F}loquet-{B}loch states on the
  surface of a topological insulator}.
\newblock \emph{\bibinfo{journal}{Science}} \textbf{\bibinfo{volume}{342}},
  \bibinfo{pages}{453--457} (\bibinfo{year}{2013}).

\bibitem{eli_deflector2}
\bibinfo{author}{Hansen, T.}
\newblock \bibinfo{title}{Laboratory directed research and development
  program}.
\newblock \emph{\bibinfo{journal}{Lawrence Berkeley National Lab}}
  (\bibinfo{year}{2011}).

\bibitem{JozwiakTOF}
\bibinfo{author}{Jozwiak, C.} \emph{et~al.}
\newblock \bibinfo{title}{A high-efficiency spin-resolved photoemission
  spectrometer combining time-of-flight spectroscopy with exchange-scattering
  polarimetry}.
\newblock \emph{\bibinfo{journal}{Review of Scientific Instruments}}
  \textbf{\bibinfo{volume}{81}}, \bibinfo{pages}{053904}
  (\bibinfo{year}{2010}).

\bibitem{deflector}
\bibinfo{author}{Wannberg, B.}
\newblock \bibinfo{title}{Analyser arrangement for particle spectrometer}.
\newblock \bibinfo{type}{U.S. Patent} \bibinfo{number}{US 9,437,408 B2},
  \bibinfo{institution}{Scienta Omicron AB} (\bibinfo{year}{2016}).

\bibitem{sunko}
\bibinfo{author}{Sunko, V.} \emph{et~al.}
\newblock \bibinfo{title}{Maximal {R}ashba-like spin splitting via
  kinetic-energy-coupled inversion-symmetry breaking}.
\newblock \emph{\bibinfo{journal}{Nature}} \textbf{\bibinfo{volume}{549}},
  \bibinfo{pages}{492--496} (\bibinfo{year}{2017}).

\bibitem{mxna}
\bibinfo{author}{Na, M.~X.} \emph{et~al.}
\newblock \bibinfo{title}{Direct determination of mode-projected
  electron-phonon coupling in the time-domain}.
\newblock \emph{\bibinfo{journal}{Preprint available at arXiv:1902.05572}}
  (\bibinfo{year}{2019}).

\end{thebibliography}

\clearpage


\section{Supplementary Materials}
\subsection{Radial Integrals and the Dipole Operator}

The main text includes a brief discussion regarding alternative approximations to the dipole operator, specifically the Fourier-representation. We expand further on the topic in this supplement. Here and throughout, we take a large simplifying assumption of a free-particle final state. Having done so, as indicated by Eq. \ref{eq:eqBFT} - \ref{eq:eqgradm1} below, the difference in the various approximations can be isolated to the form of the radial integrals: the position representation supports distinct, $l$'-dependent  final state cross sections, while the alternatives lose any such dependence. This is the defining difference, and it allows one to avoid certain final-state interference channels endemic to the momentum representation in the free-particle approximation. Such effects are explored in the figure and related discussion below. 
In consideration of the Fourier representation, we contrast:

\begin{equation}\label{eq:eqMFT_supp}
M_{FT} \propto \matrixel{ e^{i\vec{k}\cdot\vec{r}}}{\cev{\nabla}\cdot\uv{\epsilon}}{\psi_i} = i\vec{k}\cdot\uv{\epsilon}\braket{e^{i\vec{k}\cdot\vec{r}}}{\psi_i},
\end{equation}
against the position representation \cite{Jason},
\begin{equation}\label{eq:eqMpos}
M(\vec{k},\omega) \propto  \matrixel{e^{i\vec{k}\cdot\vec{r}}}{\uv\epsilon\cdot\vec{r}}{\psi_i}.
\end{equation}
In Eq. \ref{eq:eqMFT_supp}, the momentum operator has been applied to the final free electron state, reducing the matrix element to a Fourier transform of the initial state.
Further comparison of the position and Fourier representations is best done wherein the $\vec{A}\cdot\vec{k}$ and $\vec{A}\cdot\vec{r}$ matrix elements are expanded, and by making use of the properties of spherical harmonics, we can write:
\begin{equation}\label{eq:eqMk}
M(\vec{k}) = \sum_{l,m} c_{nlm}\sum_{l',\mu} B_{nl}^{l'}G_{l'-l,\mu}Y_{l'}^{-(\mu+m)*}(\Omega_k),
\end{equation}
in both cases. However, while for the Fourier transform representation,
\begin{equation}\label{eq:eqBFT}
B_{nl}^{l'} = (-i)^l\int dr r^2 j_l(kr)\phi_{nl}(r),
\end{equation}
in the position representation,
\begin{equation}\label{eq:eqBr}
B_{nl}^{l'} = (-i)^{l'}\int dr r^3 j_{l'}(kr)\phi_{nl}(r).
\end{equation}
Otherwise, the two are equivalent. It is made explicit in this form that the Fourier representation rigidly mandates radial integrals which are independent of final state angular momentum. One can perform a similar analysis of the so-called 'gradient' momentum representation  (i.e. apply the momentum operator directly to the initial state: $\vec{A}\cdot\vec{\nabla}$), where for comparison we again adopt a free-electron final state. In this case we arrive at radial integrals expressed as:
\begin{equation}\label{eq:eqgradp1}
B_{nl}^{l+1} = (-i)^{l+1}\int{dr r^2j_{l+1}(kr)(\partial_r \phi_{nl}(r) - \phi_{nl}(r)\frac{l}{r})},
\end{equation}
and,
\begin{equation}\label{eq:eqgradm1}
B_{nl}^{l-1} = (-i)^{l-1}\int{dr r^2j_{l-1}(kr)(\partial_r \phi_{nl}(r) + \phi_{nl}(r)\frac{l+1}{r})}.
\end{equation}
Integrating by parts, and exploiting properties of the spherical Bessel functions, one recovers the result of Eq. \ref{eq:eqBFT}, with $B^{l+1} = B^{l-1}$.

An essential point we want to emphasize is that while a distinction between radial integrals appears to be of critical importance in reproducing experimental data, much qualitative detail seems resilient against the precise relative cross sections. We now make this clear for a simple test case. One of the primary flaws of  $M_{FT}$ is its prediction of nodes in the photoemission intensity whenever $\vec{k}\cdot\uv{\epsilon} = 0$. For in-plane polarization then, these effects are anticipated to be quite pronounced. As is seen for the case of FeSe, under $M_{FT}$, intensity should vanish for $k_y=0$ in panel (a) of Fig. 1 of the main text. On the contrary, this is where intensity is in fact largest. Taking this as a motivating example, we construct two circular Fermi surface contours defined by helical orbital texture, as anticipated for hole-doped Fe-based superconductors \cite{watson,Day}. To make these claims more generic, we perform a similar analysis for analogous wavefunctions composed of $p$, rather than $d$ orbitals. Explicitly, we use the following wavefunctions: 
\\
\begin{center}
\begin{tabular}{ |c| }
\hline
Fe: $Z = 26 $ $n=3$ $ l=2$ \\
\hline
$\psi_{outer} = cos(\phi)d_{yz} - sin (\phi)d_{xz}$  \\
$\psi_{inner} = sin(\phi)d_{yz} + cos(\phi)d_{xz}$ \\
\hline
 C: $Z = 6$ $n=2$  $ l=1$\\
 \hline
 $\psi_{outer} = cos(\phi) p_y - sin(\phi) p_x$\\
 $\psi_{inner} = sin(\phi) p_y + cos(\phi) p_x$\\
\hline

\end{tabular}
\end{center}
\vspace{20pt}
The subscripts inner (outer) refer to the states with smaller (larger) $k_F$. In addition, $\phi=tan^{-1}(k_y/k_x)$. The orbital projections are indicated schematically in the inset of panel (a) of the figure. As elsewhere, we assume a free electron final state in our calculation of the ARPES matrix element. 

In the figure, we compare and contrast the effect of identical ($B^{l+1} = B^{l-1}$) and distinct ($B^{l+1} \neq B^{l-1}$) radial integrals on this model system. For $\uv{\epsilon} = [0,1,0]$, $M_{FT}$ predicts a node in the photoemission intensity along $k_y=0$. A reasonable metric then for comparison of $B^{l-1}/B^{l+1}$ is the ratio of maximum intensities along the $y$ axis ($I_y$ where $k_x=0$) and $x$ axis ($I_x$, $k_y=0$), labeled here as $I_x/I_y$. The specific momentum points used are indicated by blue crosses in panel (b) of the figure. For $M_{FT}$, we expect $I_x/I_y$ to be zero. The blue (yellow) curve in (a) corresponds to this intensity ratio $I_x/I_y$ for the $l = 2$ ($l=1$) states. Constant energy contours (i.e. “Fermi surfaces”) at representative points from (a) are plotted in panels (b-e). The photoemission pattern is largely insensitive to the particular ratio $B^{l-1}/B^{l+1}$ away from the singularity at $B^{l-1}/B^{l+1} = 1$.
\begin{figure}[t!]
\includegraphics[width=\columnwidth]{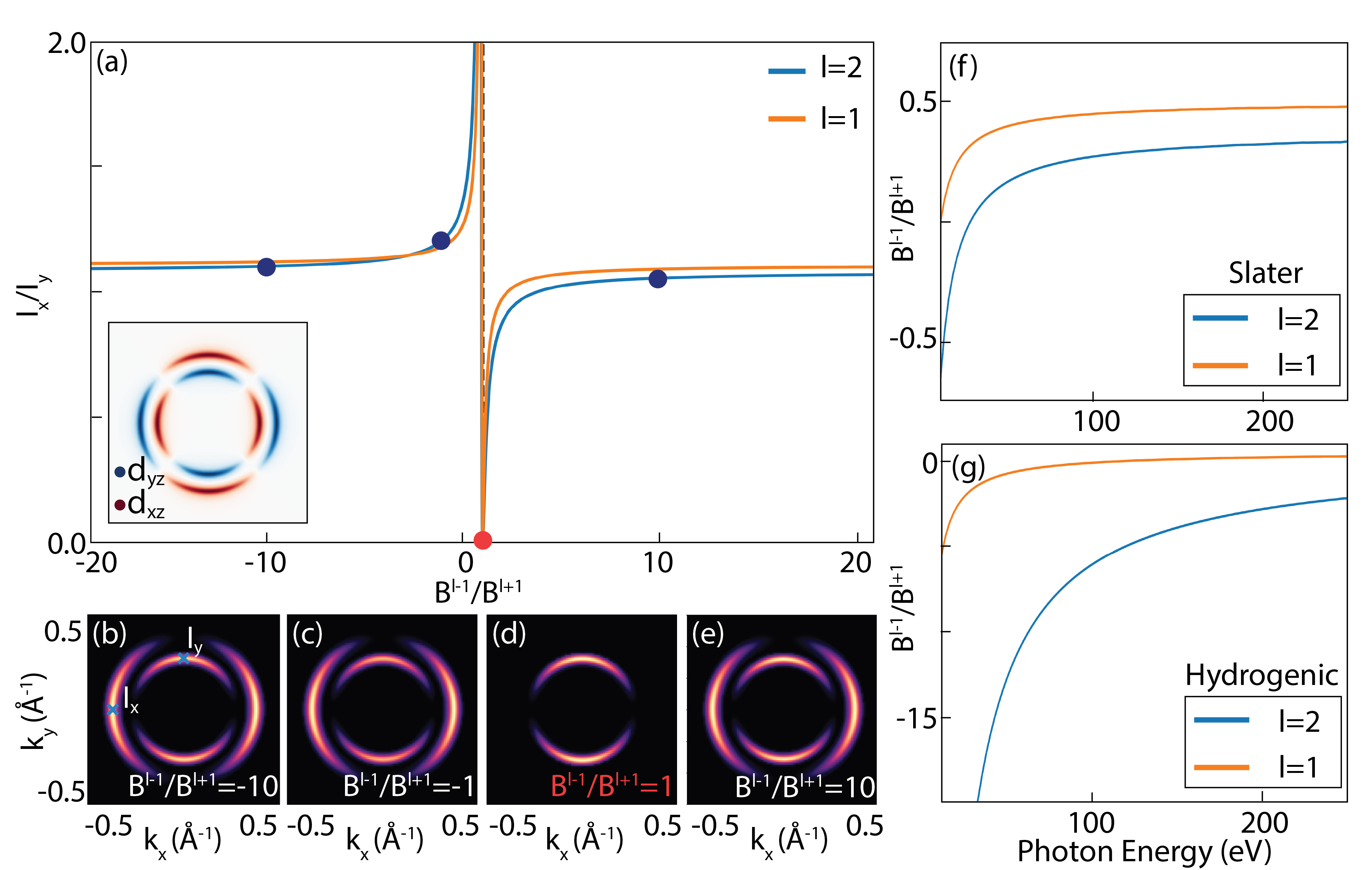}
\caption{Radial Integral Dependence of Predicted ARPES pattern. In (a), photoemission intensity ratio along $k_y$ and $k_x$ axes ($I_x/I_y$), as indicated by blue crosses in (b). This ratio is plotted as a function of $B^{l-1}/B^{l+1}$, for $l=1$ (yellow) and $l=2$ (blue) initial states. In (b-e), representative intensity maps used in the calculation of (a) are displayed, corresponding to the circles in (a). Panels (f,g) illustrate the range of $B^{l-1}/B^{l+1}$ predicted for different photon energies,for $l=2$ and $l=1$ initial states. This is done for choice of Slater (f) and hydrogenic (g) orbitals.}
\end{figure}

As illustrated in panels (b-e), the node at $k_y=0$ is observed only at this singularity, i.e. where $M_{FT}$ is used. Such a node is not observed away from the value $B^{l-1}/B^{l+1}=1$. The failure of $M_{FT}$ goes beyond this erroneous node. In addition, intensity from the outer state is vanishing along the entire circular contour. By contrast, the patterns in (b,c,e) are more representative of experimental data. Furthermore, the angular intensity pattern is qualitatively, and nearly quantitatively, identical over most of the rest of the domain of (a). As one naively expects, $\uv{\epsilon} = [0,1,0]$ is primarily sensitive to $d_{yz}$ orbitals for small $k_{||}$. This is not strictly true in panel (d), which is calculated in the Fourier representation, as the $d_{yz}$ component of $\psi_{outer}$ does not generate any intensity. Calculations for $l=1$ initial states perform similarly, indicated by the yellow curve in panel (a). Evidently, it is essential that these integrals be inequivalent for different $l$' in order to recover something akin to the experimental results. While the Fourier representation rigidly imposes $B^{l-1}/B^{l+1} = 1$, the position representation, as for more sophisticated approximations to the final state, support ratios away from this value.

Despite the broad insensitivity to $B^{l-1}/B^{l+1} $, it is instructive to indicate where calculations done in the position representation land on the axis of (a). To provide a connection to actual radial integrals we have then plotted, in panels (f) and (g) the photon energy dependence of radial integrals calculated in the position representation. This is done here for Slater (f) and hydrogenic (g) orbitals. The two curves reflect an $l=1$ (yellow, Carbon, $Z=6$) and $l=2$ (blue, Iron, $Z=26$) initial state. While $B^{l-1}/B^{l+1}$ varies significantly with photon energy and choice of initial state radial function, the singularity at $B^{l-1}/B^{l+1} = 1$ is not crossed within this broad domain. 

As suggested by the distinctions between panels (f) and (g), in relation to a typical parameterized tight-binding model treatment of the electronic structure used in the modelling of ARPES data, the initial state wavefunctions represent to some extent another degree of freedom in the treatment of the relevant problem. The asymptotic behaviour of these same panels indicate that the position representation reflects an approximation to the photoemission event which converges towards the momentum representation in the high-energy limit. In the low-energy limit where experiments are typically conducted, this framework provides good agreement with experiment through the ability to parameterize the radial integrals with distinct final-state angular momentum cross sections. It is instructive to note that although the plane-wave approximation is a large simplification, the $l$'-dependent cross sections of the position representation facilitate a natural connection to proper final states, where scattering phase shifts mandate different cross sections for each angular momentum final state channel. By contrast, the Fourier representation is not amenable to calculations wherein proper final states are introduced, as $\nabla\psi_f\neq i\vec{k}\psi_f$  and $\braket{\psi_f}{\psi_i}\neq \psi_i(\vec{k})$. Future developments, employing for example DFT-derived initial and final state wavefunctions, will allow for a more rigourously motivated description of the relevant processes.

\end{document}